\documentclass[a4paper,traditabstract,longauth]{aa}

\usepackage{graphicx}
\usepackage{txfonts}
\usepackage{natbib}
\usepackage{longtable,lscape}
\usepackage{amssymb}
\usepackage{mathrsfs}
\usepackage{amsfonts}
\usepackage{url}
\usepackage{color}
\usepackage[colorlinks, citecolor=blue]{hyperref}

\usepackage{fixltx2e}

\newcommand{\zphot}{z_{\rm phot}}

\def\setsymbol#1#2{\expandafter\def\csname #1\endcsname{#2}}
\def\getsymbol#1{\csname #1\endcsname}

\def\Planck{\textit{Planck}}

\def\all2013resultspapers{\nocite{planck2013-p01, planck2013-p02, planck2013-p02a, planck2013-p02d, planck2013-p02b, planck2013-p03, planck2013-p03c, planck2013-p03f, planck2013-p03d, planck2013-p03e, planck2013-p01a, planck2013-p06, planck2013-p03a, planck2013-pip88, planck2013-p08, planck2013-p11, planck2013-p12, planck2013-p13, planck2013-p14, planck2013-p15, planck2013-p05b, planck2013-p17, planck2013-p09, planck2013-p09a, planck2013-p20, planck2013-p19, planck2013-pipaberration, planck2013-p05, planck2013-p05a, planck2013-pip56, planck2013-p06b}}

\newbox\tablebox    \newdimen\tablewidth
\def\leaderfil{\leaders\hbox to 5pt{\hss.\hss}\hfil}
\def\endPlancktable{\tablewidth=\columnwidth 
    $$\hss\copy\tablebox\hss$$
    \vskip-\lastskip\vskip -2pt}
\def\endPlancktablewide{\tablewidth=\textwidth 
    $$\hss\copy\tablebox\hss$$
    \vskip-\lastskip\vskip -2pt}
\def\tablenote#1 #2\par{\begingroup \parindent=0.8em
    \abovedisplayshortskip=0pt\belowdisplayshortskip=0pt
    \noindent
    $$\hss\vbox{\hsize\tablewidth \hangindent=\parindent \hangafter=1 \noindent
    \hbox to \parindent{$^#1$\hss}\strut#2\strut\par}\hss$$
    \endgroup}
\def\doubleline{\vskip 3pt\hrule \vskip 1.5pt \hrule \vskip 5pt}

\def\L2{\ifmmode L_2\else $L_2$\fi}

\def\DeltaT{\ifmmode \Delta T\else $\Delta T$\fi}
\def\deltat{\ifmmode \Delta t\else $\Delta t$\fi}
\def\fknee{\ifmmode f_{\rm knee}\else $f_{\rm knee}$\fi}
\def\Fmax{\ifmmode F_{\rm max}\else $F_{\rm max}$\fi}
\def\solar{\ifmmode{\rm M}_{\mathord\odot}\else${\rm M}_{\mathord\odot}$\fi}
\def\Msolar{\ifmmode{\rm M}_{\mathord\odot}\else${\rm M}_{\mathord\odot}$\fi}
\def\Lsolar{\ifmmode{\rm L}_{\mathord\odot}\else${\rm L}_{\mathord\odot}$\fi}

\def\inv{\ifmmode^{-1}\else$^{-1}$\fi}
\def\mo{\ifmmode^{-1}\else$^{-1}$\fi}
\def\sup#1{\ifmmode ^{\rm #1}\else $^{\rm #1}$\fi}
\def\expo#1{\ifmmode \times 10^{#1}\else $\times 10^{#1}$\fi}
\def\,{\thinspace}
\def\lsim{\mathrel{\raise .4ex\hbox{\rlap{$<$}\lower 1.2ex\hbox{$\sim$}}}}
\def\gsim{\mathrel{\raise .4ex\hbox{\rlap{$>$}\lower 1.2ex\hbox{$\sim$}}}}

\def\simprop{\mathrel{\raise .4ex\hbox{\rlap{$\propto$}\lower 1.2ex\hbox{$\sim$}}}}
\def\deg{\ifmmode^\circ\else$^\circ$\fi}
\def\pdeg{\ifmmode $\setbox0=\hbox{$^{\circ}$}\rlap{\hskip.11\wd0 .}$^{\circ}
          \else \setbox0=\hbox{$^{\circ}$}\rlap{\hskip.11\wd0 .}$^{\circ}$\fi}
\def\arcs{\ifmmode {^{\scriptstyle\prime\prime}}
          \else $^{\scriptstyle\prime\prime}$\fi}
\def\arcm{\ifmmode {^{\scriptstyle\prime}}
          \else $^{\scriptstyle\prime}$\fi}
\newdimen\sa  \newdimen\sb
\def\parcs{\sa=.07em \sb=.03em
     \ifmmode \hbox{\rlap{.}}^{\scriptstyle\prime\kern -\sb\prime}\hbox{\kern -\sa}
     \else \rlap{.}$^{\scriptstyle\prime\kern -\sb\prime}$\kern -\sa\fi}
\def\parcm{\sa=.08em \sb=.03em
     \ifmmode \hbox{\rlap{.}\kern\sa}^{\scriptstyle\prime}\hbox{\kern-\sb}
     \else \rlap{.}\kern\sa$^{\scriptstyle\prime}$\kern-\sb\fi}
\def\ra[#1 #2 #3.#4]{#1\sup{h}#2\sup{m}#3\sup{s}\llap.#4}
\def\dec[#1 #2 #3.#4]{#1\deg#2\arcm#3\arcs\llap.#4}
\def\deco[#1 #2 #3]{#1\deg#2\arcm#3\arcs}
\def\rra[#1 #2]{#1\sup{h}#2\sup{m}}

\def\dots{\relax\ifmmode \ldots\else $\ldots$\fi}
\def\WHzsr{\ifmmode $W\,Hz\mo\,sr\mo$\else W\,Hz\mo\,sr\mo\fi}
\def\mHz{\ifmmode $\,mHz$\else \,mHz\fi}
\def\GHz{\ifmmode $\,GHz$\else \,GHz\fi}
\def\mKs{\ifmmode $\,mK\,s$^{1/2}\else \,mK\,s$^{1/2}$\fi}
\def\muKs{\ifmmode \,\mu$K\,s$^{1/2}\else \,$\mu$K\,s$^{1/2}$\fi}
\def\muKRJs{\ifmmode \,\mu$K$_{\rm RJ}$\,s$^{1/2}\else \,$\mu$K$_{\rm RJ}$\,s$^{1/2}$\fi}
\def\muKHz{\ifmmode \,\mu$K\,Hz$^{-1/2}\else \,$\mu$K\,Hz$^{-1/2}$\fi}
\def\MJysr{\ifmmode \,$MJy\,sr\mo$\else \,MJy\,sr\mo\fi}
\def\MJysrmK{\ifmmode \,$MJy\,sr\mo$\,mK$_{\rm CMB}\mo\else \,MJy\,sr\mo\,mK$_{\rm CMB}\mo$\fi}
\def\microns{\ifmmode \,\mu$m$\else \,$\mu$m\fi}

\def\muK{\ifmmode \,\mu$K$\else \,$\mu$\hbox{K}\fi}
\def\microK{\ifmmode \,\mu$K$\else \,$\mu$\hbox{K}\fi}
\def\muW{\ifmmode \,\mu$W$\else \,$\mu$\hbox{W}\fi}
\def\kms{\ifmmode $\,km\,s$^{-1}\else \,km\,s$^{-1}$\fi}
\def\kmsMpc{\ifmmode $\,\kms\,Mpc\mo$\else \,\kms\,Mpc\mo\fi}
\setsymbol{LFI:center:frequency:70GHz:units}{70.3\,GHz}
\setsymbol{LFI:center:frequency:44GHz:units}{44.1\,GHz}
\setsymbol{LFI:center:frequency:30GHz:units}{28.5\,GHz}

\setsymbol{LFI:center:frequency:70GHz}{70.3}
\setsymbol{LFI:center:frequency:44GHz}{44.1}
\setsymbol{LFI:center:frequency:30GHz}{28.5}

\setsymbol{LFI:center:frequency:LFI18:Rad:M:units}{71.7\GHz}
\setsymbol{LFI:center:frequency:LFI19:Rad:M:units}{67.5\GHz}
\setsymbol{LFI:center:frequency:LFI20:Rad:M:units}{69.2\GHz}
\setsymbol{LFI:center:frequency:LFI21:Rad:M:units}{70.4\GHz}
\setsymbol{LFI:center:frequency:LFI22:Rad:M:units}{71.5\GHz}
\setsymbol{LFI:center:frequency:LFI23:Rad:M:units}{70.8\GHz}
\setsymbol{LFI:center:frequency:LFI24:Rad:M:units}{44.4\GHz}
\setsymbol{LFI:center:frequency:LFI25:Rad:M:units}{44.0\GHz}
\setsymbol{LFI:center:frequency:LFI26:Rad:M:units}{43.9\GHz}
\setsymbol{LFI:center:frequency:LFI27:Rad:M:units}{28.3\GHz}
\setsymbol{LFI:center:frequency:LFI28:Rad:M:units}{28.8\GHz}
\setsymbol{LFI:center:frequency:LFI18:Rad:S:units}{70.1\GHz}
\setsymbol{LFI:center:frequency:LFI19:Rad:S:units}{69.6\GHz}
\setsymbol{LFI:center:frequency:LFI20:Rad:S:units}{69.5\GHz}
\setsymbol{LFI:center:frequency:LFI21:Rad:S:units}{69.5\GHz}
\setsymbol{LFI:center:frequency:LFI22:Rad:S:units}{72.8\GHz}
\setsymbol{LFI:center:frequency:LFI23:Rad:S:units}{71.3\GHz}
\setsymbol{LFI:center:frequency:LFI24:Rad:S:units}{44.1\GHz}
\setsymbol{LFI:center:frequency:LFI25:Rad:S:units}{44.1\GHz}
\setsymbol{LFI:center:frequency:LFI26:Rad:S:units}{44.1\GHz}
\setsymbol{LFI:center:frequency:LFI27:Rad:S:units}{28.5\GHz}
\setsymbol{LFI:center:frequency:LFI28:Rad:S:units}{28.2\GHz}

\setsymbol{LFI:center:frequency:LFI18:Rad:M}{71.7}
\setsymbol{LFI:center:frequency:LFI19:Rad:M}{67.5}
\setsymbol{LFI:center:frequency:LFI20:Rad:M}{69.2}
\setsymbol{LFI:center:frequency:LFI21:Rad:M}{70.4}
\setsymbol{LFI:center:frequency:LFI22:Rad:M}{71.5}
\setsymbol{LFI:center:frequency:LFI23:Rad:M}{70.8}
\setsymbol{LFI:center:frequency:LFI24:Rad:M}{44.4}
\setsymbol{LFI:center:frequency:LFI25:Rad:M}{44.0}
\setsymbol{LFI:center:frequency:LFI26:Rad:M}{43.9}
\setsymbol{LFI:center:frequency:LFI27:Rad:M}{28.3}
\setsymbol{LFI:center:frequency:LFI28:Rad:M}{28.8}
\setsymbol{LFI:center:frequency:LFI18:Rad:S}{70.1}
\setsymbol{LFI:center:frequency:LFI19:Rad:S}{69.6}
\setsymbol{LFI:center:frequency:LFI20:Rad:S}{69.5}
\setsymbol{LFI:center:frequency:LFI21:Rad:S}{69.5}
\setsymbol{LFI:center:frequency:LFI22:Rad:S}{72.8}
\setsymbol{LFI:center:frequency:LFI23:Rad:S}{71.3}
\setsymbol{LFI:center:frequency:LFI24:Rad:S}{44.1}
\setsymbol{LFI:center:frequency:LFI25:Rad:S}{44.1}
\setsymbol{LFI:center:frequency:LFI26:Rad:S}{44.1}
\setsymbol{LFI:center:frequency:LFI27:Rad:S}{28.5}
\setsymbol{LFI:center:frequency:LFI28:Rad:S}{28.2}

\setsymbol{LFI:white:noise:sensitivity:70GHz:units}{134.7\muKs}
\setsymbol{LFI:white:noise:sensitivity:44GHz:units}{164.7\muKs}
\setsymbol{LFI:white:noise:sensitivity:30GHz:units}{143.4\muKs}

\setsymbol{LFI:white:noise:sensitivity:70GHz}{134.7}
\setsymbol{LFI:white:noise:sensitivity:44GHz}{164.7}
\setsymbol{LFI:white:noise:sensitivity:30GHz}{143.4}

\setsymbol{LFI:white:noise:sensitivity:LFI18:Rad:M:units}{512.0\muKs}
\setsymbol{LFI:white:noise:sensitivity:LFI19:Rad:M:units}{581.4\muKs}
\setsymbol{LFI:white:noise:sensitivity:LFI20:Rad:M:units}{590.8\muKs}
\setsymbol{LFI:white:noise:sensitivity:LFI21:Rad:M:units}{455.2\muKs}
\setsymbol{LFI:white:noise:sensitivity:LFI22:Rad:M:units}{492.0\muKs}
\setsymbol{LFI:white:noise:sensitivity:LFI23:Rad:M:units}{507.7\muKs}
\setsymbol{LFI:white:noise:sensitivity:LFI24:Rad:M:units}{462.2\muKs}
\setsymbol{LFI:white:noise:sensitivity:LFI25:Rad:M:units}{413.6\muKs}
\setsymbol{LFI:white:noise:sensitivity:LFI26:Rad:M:units}{478.6\muKs}
\setsymbol{LFI:white:noise:sensitivity:LFI27:Rad:M:units}{277.7\muKs}
\setsymbol{LFI:white:noise:sensitivity:LFI28:Rad:M:units}{312.3\muKs}
\setsymbol{LFI:white:noise:sensitivity:LFI18:Rad:S:units}{465.7\muKs}
\setsymbol{LFI:white:noise:sensitivity:LFI19:Rad:S:units}{555.6\muKs}
\setsymbol{LFI:white:noise:sensitivity:LFI20:Rad:S:units}{623.2\muKs}
\setsymbol{LFI:white:noise:sensitivity:LFI21:Rad:S:units}{564.1\muKs}
\setsymbol{LFI:white:noise:sensitivity:LFI22:Rad:S:units}{534.4\muKs}
\setsymbol{LFI:white:noise:sensitivity:LFI23:Rad:S:units}{542.4\muKs}
\setsymbol{LFI:white:noise:sensitivity:LFI24:Rad:S:units}{399.2\muKs}
\setsymbol{LFI:white:noise:sensitivity:LFI25:Rad:S:units}{392.6\muKs}
\setsymbol{LFI:white:noise:sensitivity:LFI26:Rad:S:units}{418.6\muKs}
\setsymbol{LFI:white:noise:sensitivity:LFI27:Rad:S:units}{302.9\muKs}
\setsymbol{LFI:white:noise:sensitivity:LFI28:Rad:S:units}{285.3\muKs}

\setsymbol{LFI:white:noise:sensitivity:LFI18:Rad:M}{512.0}
\setsymbol{LFI:white:noise:sensitivity:LFI19:Rad:M}{581.4}
\setsymbol{LFI:white:noise:sensitivity:LFI20:Rad:M}{590.8}
\setsymbol{LFI:white:noise:sensitivity:LFI21:Rad:M}{455.2}
\setsymbol{LFI:white:noise:sensitivity:LFI22:Rad:M}{492.0}
\setsymbol{LFI:white:noise:sensitivity:LFI23:Rad:M}{507.7}
\setsymbol{LFI:white:noise:sensitivity:LFI24:Rad:M}{462.2}
\setsymbol{LFI:white:noise:sensitivity:LFI25:Rad:M}{413.6}
\setsymbol{LFI:white:noise:sensitivity:LFI26:Rad:M}{478.6}
\setsymbol{LFI:white:noise:sensitivity:LFI27:Rad:M}{277.7}
\setsymbol{LFI:white:noise:sensitivity:LFI28:Rad:M}{312.3}
\setsymbol{LFI:white:noise:sensitivity:LFI18:Rad:S}{465.7}
\setsymbol{LFI:white:noise:sensitivity:LFI19:Rad:S}{555.6}
\setsymbol{LFI:white:noise:sensitivity:LFI20:Rad:S}{623.2}
\setsymbol{LFI:white:noise:sensitivity:LFI21:Rad:S}{564.1}
\setsymbol{LFI:white:noise:sensitivity:LFI22:Rad:S}{534.4}
\setsymbol{LFI:white:noise:sensitivity:LFI23:Rad:S}{542.4}
\setsymbol{LFI:white:noise:sensitivity:LFI24:Rad:S}{399.2}
\setsymbol{LFI:white:noise:sensitivity:LFI25:Rad:S}{392.6}
\setsymbol{LFI:white:noise:sensitivity:LFI26:Rad:S}{418.6}
\setsymbol{LFI:white:noise:sensitivity:LFI27:Rad:S}{302.9}
\setsymbol{LFI:white:noise:sensitivity:LFI28:Rad:S}{285.3}

\setsymbol{LFI:knee:frequency:70GHz:units}{29.5\mHz}
\setsymbol{LFI:knee:frequency:44GHz:units}{56.2\mHz}
\setsymbol{LFI:knee:frequency:30GHz:units}{113.7\mHz}

\setsymbol{LFI:knee:frequency:70GHz}{29.5}
\setsymbol{LFI:knee:frequency:44GHz}{56.2}
\setsymbol{LFI:knee:frequency:30GHz}{113.7}

\setsymbol{LFI:knee:frequency:LFI18:Rad:M:units}{16.3\mHz}
\setsymbol{LFI:knee:frequency:LFI19:Rad:M:units}{15.1\mHz}
\setsymbol{LFI:knee:frequency:LFI20:Rad:M:units}{18.7\mHz}
\setsymbol{LFI:knee:frequency:LFI21:Rad:M:units}{37.2\mHz}
\setsymbol{LFI:knee:frequency:LFI22:Rad:M:units}{12.7\mHz}
\setsymbol{LFI:knee:frequency:LFI23:Rad:M:units}{34.6\mHz}
\setsymbol{LFI:knee:frequency:LFI24:Rad:M:units}{46.2\mHz}
\setsymbol{LFI:knee:frequency:LFI25:Rad:M:units}{24.9\mHz}
\setsymbol{LFI:knee:frequency:LFI26:Rad:M:units}{67.6\mHz}
\setsymbol{LFI:knee:frequency:LFI27:Rad:M:units}{187.4\mHz}
\setsymbol{LFI:knee:frequency:LFI28:Rad:M:units}{122.2\mHz}
\setsymbol{LFI:knee:frequency:LFI18:Rad:S:units}{17.7\mHz}
\setsymbol{LFI:knee:frequency:LFI19:Rad:S:units}{22.0\mHz}
\setsymbol{LFI:knee:frequency:LFI20:Rad:S:units}{8.7\mHz}
\setsymbol{LFI:knee:frequency:LFI21:Rad:S:units}{25.9\mHz}
\setsymbol{LFI:knee:frequency:LFI22:Rad:S:units}{15.8\mHz}
\setsymbol{LFI:knee:frequency:LFI23:Rad:S:units}{129.8\mHz}
\setsymbol{LFI:knee:frequency:LFI24:Rad:S:units}{100.9\mHz}
\setsymbol{LFI:knee:frequency:LFI25:Rad:S:units}{38.9\mHz}
\setsymbol{LFI:knee:frequency:LFI26:Rad:S:units}{58.9\mHz}
\setsymbol{LFI:knee:frequency:LFI27:Rad:S:units}{104.4\mHz}
\setsymbol{LFI:knee:frequency:LFI28:Rad:S:units}{40.7\mHz}

\setsymbol{LFI:knee:frequency:LFI18:Rad:M}{16.3}
\setsymbol{LFI:knee:frequency:LFI19:Rad:M}{15.1}
\setsymbol{LFI:knee:frequency:LFI20:Rad:M}{18.7}
\setsymbol{LFI:knee:frequency:LFI21:Rad:M}{37.2}
\setsymbol{LFI:knee:frequency:LFI22:Rad:M}{12.7}
\setsymbol{LFI:knee:frequency:LFI23:Rad:M}{34.6}
\setsymbol{LFI:knee:frequency:LFI24:Rad:M}{46.2}
\setsymbol{LFI:knee:frequency:LFI25:Rad:M}{24.9}
\setsymbol{LFI:knee:frequency:LFI26:Rad:M}{67.6}
\setsymbol{LFI:knee:frequency:LFI27:Rad:M}{187.4}
\setsymbol{LFI:knee:frequency:LFI28:Rad:M}{122.2}
\setsymbol{LFI:knee:frequency:LFI18:Rad:S}{17.7}
\setsymbol{LFI:knee:frequency:LFI19:Rad:S}{22.0}
\setsymbol{LFI:knee:frequency:LFI20:Rad:S}{8.7}
\setsymbol{LFI:knee:frequency:LFI21:Rad:S}{25.9}
\setsymbol{LFI:knee:frequency:LFI22:Rad:S}{15.8}
\setsymbol{LFI:knee:frequency:LFI23:Rad:S}{129.8}
\setsymbol{LFI:knee:frequency:LFI24:Rad:S}{100.9}
\setsymbol{LFI:knee:frequency:LFI25:Rad:S}{38.9}
\setsymbol{LFI:knee:frequency:LFI26:Rad:S}{58.9}
\setsymbol{LFI:knee:frequency:LFI27:Rad:S}{104.4}
\setsymbol{LFI:knee:frequency:LFI28:Rad:S}{40.7}

\setsymbol{LFI:slope:70GHz:units}{$-1.03$\mHz}
\setsymbol{LFI:slope:44GHz:units}{$-0.89$\mHz}
\setsymbol{LFI:slope:30GHz:units}{$-0.87$\mHz}

\setsymbol{LFI:slope:70GHz}{$-1.03$}
\setsymbol{LFI:slope:44GHz}{$-0.89$}
\setsymbol{LFI:slope:30GHz}{$-0.87$}

\setsymbol{LFI:slope:LFI18:Rad:M:units}{$-1.04$\mHz}
\setsymbol{LFI:slope:LFI19:Rad:M:units}{$-1.09$\mHz}
\setsymbol{LFI:slope:LFI20:Rad:M:units}{$-0.69$\mHz}
\setsymbol{LFI:slope:LFI21:Rad:M:units}{$-1.56$\mHz}
\setsymbol{LFI:slope:LFI22:Rad:M:units}{$-1.01$\mHz}
\setsymbol{LFI:slope:LFI23:Rad:M:units}{$-0.96$\mHz}
\setsymbol{LFI:slope:LFI24:Rad:M:units}{$-0.83$\mHz}
\setsymbol{LFI:slope:LFI25:Rad:M:units}{$-0.91$\mHz}
\setsymbol{LFI:slope:LFI26:Rad:M:units}{$-0.95$\mHz}
\setsymbol{LFI:slope:LFI27:Rad:M:units}{$-0.87$\mHz}
\setsymbol{LFI:slope:LFI28:Rad:M:units}{$-0.88$\mHz}
\setsymbol{LFI:slope:LFI18:Rad:S:units}{$-1.15$\mHz}
\setsymbol{LFI:slope:LFI19:Rad:S:units}{$-1.00$\mHz}
\setsymbol{LFI:slope:LFI20:Rad:S:units}{$-0.95$\mHz}
\setsymbol{LFI:slope:LFI21:Rad:S:units}{$-0.92$\mHz}
\setsymbol{LFI:slope:LFI22:Rad:S:units}{$-1.01$\mHz}
\setsymbol{LFI:slope:LFI23:Rad:S:units}{$-0.95$\mHz}
\setsymbol{LFI:slope:LFI24:Rad:S:units}{$-0.73$\mHz}
\setsymbol{LFI:slope:LFI25:Rad:S:units}{$-1.16$\mHz}
\setsymbol{LFI:slope:LFI26:Rad:S:units}{$-0.79$\mHz}
\setsymbol{LFI:slope:LFI27:Rad:S:units}{$-0.82$\mHz}
\setsymbol{LFI:slope:LFI28:Rad:S:units}{$-0.91$\mHz}

\setsymbol{LFI:slope:LFI18:Rad:M}{$-1.04$}
\setsymbol{LFI:slope:LFI19:Rad:M}{$-1.09$}
\setsymbol{LFI:slope:LFI20:Rad:M}{$-0.69$}
\setsymbol{LFI:slope:LFI21:Rad:M}{$-1.56$}
\setsymbol{LFI:slope:LFI22:Rad:M}{$-1.01$}
\setsymbol{LFI:slope:LFI23:Rad:M}{$-0.96$}
\setsymbol{LFI:slope:LFI24:Rad:M}{$-0.83$}
\setsymbol{LFI:slope:LFI25:Rad:M}{$-0.91$}
\setsymbol{LFI:slope:LFI26:Rad:M}{$-0.95$}
\setsymbol{LFI:slope:LFI27:Rad:M}{$-0.87$}
\setsymbol{LFI:slope:LFI28:Rad:M}{$-0.88$}
\setsymbol{LFI:slope:LFI18:Rad:S}{$-1.15$}
\setsymbol{LFI:slope:LFI19:Rad:S}{$-1.00$}
\setsymbol{LFI:slope:LFI20:Rad:S}{$-0.95$}
\setsymbol{LFI:slope:LFI21:Rad:S}{$-0.92$}
\setsymbol{LFI:slope:LFI22:Rad:S}{$-1.01$}
\setsymbol{LFI:slope:LFI23:Rad:S}{$-0.95$}
\setsymbol{LFI:slope:LFI24:Rad:S}{$-0.73$}
\setsymbol{LFI:slope:LFI25:Rad:S}{$-1.16$}
\setsymbol{LFI:slope:LFI26:Rad:S}{$-0.79$}
\setsymbol{LFI:slope:LFI27:Rad:S}{$-0.82$}
\setsymbol{LFI:slope:LFI28:Rad:S}{$-0.91$}

\setsymbol{LFI:FWHM:70GHz:units}{13\parcm01}
\setsymbol{LFI:FWHM:44GHz:units}{27\parcm92}
\setsymbol{LFI:FWHM:30GHz:units}{32\parcm65}

\setsymbol{LFI:FWHM:70GHz}{13.01}
\setsymbol{LFI:FWHM:44GHz}{27.92}
\setsymbol{LFI:FWHM:30GHz}{32.65}

\setsymbol{LFI:FWHM:LFI18:units}{13\parcm39}
\setsymbol{LFI:FWHM:LFI19:units}{13\parcm01}
\setsymbol{LFI:FWHM:LFI20:units}{12\parcm75}
\setsymbol{LFI:FWHM:LFI21:units}{12\parcm74}
\setsymbol{LFI:FWHM:LFI22:units}{12\parcm87}
\setsymbol{LFI:FWHM:LFI23:units}{13\parcm27}
\setsymbol{LFI:FWHM:LFI24:units}{22\parcm98}
\setsymbol{LFI:FWHM:LFI25:units}{30\parcm46}
\setsymbol{LFI:FWHM:LFI26:units}{30\parcm31}
\setsymbol{LFI:FWHM:LFI27:units}{32\parcm65}
\setsymbol{LFI:FWHM:LFI28:units}{32\parcm66}

\setsymbol{LFI:FWHM:LFI18}{13.39}
\setsymbol{LFI:FWHM:LFI19}{13.01}
\setsymbol{LFI:FWHM:LFI20}{12.75}
\setsymbol{LFI:FWHM:LFI21}{12.74}
\setsymbol{LFI:FWHM:LFI22}{12.87}
\setsymbol{LFI:FWHM:LFI23}{13.27}
\setsymbol{LFI:FWHM:LFI24}{22.98}
\setsymbol{LFI:FWHM:LFI25}{30.46}
\setsymbol{LFI:FWHM:LFI26}{30.31}
\setsymbol{LFI:FWHM:LFI27}{32.65}
\setsymbol{LFI:FWHM:LFI28}{32.66}

\setsymbol{LFI:FWHM:uncertainty:LFI18:units}{0.170\arcm}
\setsymbol{LFI:FWHM:uncertainty:LFI19:units}{0.174\arcm}
\setsymbol{LFI:FWHM:uncertainty:LFI20:units}{0.170\arcm}
\setsymbol{LFI:FWHM:uncertainty:LFI21:units}{0.156\arcm}
\setsymbol{LFI:FWHM:uncertainty:LFI22:units}{0.164\arcm}
\setsymbol{LFI:FWHM:uncertainty:LFI23:units}{0.171\arcm}
\setsymbol{LFI:FWHM:uncertainty:LFI24:units}{0.652\arcm}
\setsymbol{LFI:FWHM:uncertainty:LFI25:units}{1.075\arcm}
\setsymbol{LFI:FWHM:uncertainty:LFI26:units}{1.131\arcm}
\setsymbol{LFI:FWHM:uncertainty:LFI27:units}{1.266\arcm}
\setsymbol{LFI:FWHM:uncertainty:LFI28:units}{1.287\arcm}

\setsymbol{LFI:FWHM:uncertainty:LFI18}{0.170}
\setsymbol{LFI:FWHM:uncertainty:LFI19}{0.174}
\setsymbol{LFI:FWHM:uncertainty:LFI20}{0.170}
\setsymbol{LFI:FWHM:uncertainty:LFI21}{0.156}
\setsymbol{LFI:FWHM:uncertainty:LFI22}{0.164}
\setsymbol{LFI:FWHM:uncertainty:LFI23}{0.171}
\setsymbol{LFI:FWHM:uncertainty:LFI24}{0.652}
\setsymbol{LFI:FWHM:uncertainty:LFI25}{1.075}
\setsymbol{LFI:FWHM:uncertainty:LFI26}{1.131}
\setsymbol{LFI:FWHM:uncertainty:LFI27}{1.266}
\setsymbol{LFI:FWHM:uncertainty:LFI28}{1.287}

\setsymbol{HFI:center:frequency:100GHz:units}{100\,GHz}
\setsymbol{HFI:center:frequency:143GHz:units}{143\,GHz}
\setsymbol{HFI:center:frequency:217GHz:units}{217\,GHz}
\setsymbol{HFI:center:frequency:353GHz:units}{353\,GHz}
\setsymbol{HFI:center:frequency:545GHz:units}{545\,GHz}
\setsymbol{HFI:center:frequency:857GHz:units}{857\,GHz}

\setsymbol{HFI:center:frequency:100GHz}{100}
\setsymbol{HFI:center:frequency:143GHz}{143}
\setsymbol{HFI:center:frequency:217GHz}{217}
\setsymbol{HFI:center:frequency:353GHz}{353}
\setsymbol{HFI:center:frequency:545GHz}{545}
\setsymbol{HFI:center:frequency:857GHz}{857}

\setsymbol{HFI:Ndetectors:100GHz}{8}
\setsymbol{HFI:Ndetectors:143GHz}{11}
\setsymbol{HFI:Ndetectors:217GHz}{12}
\setsymbol{HFI:Ndetectors:353GHz}{12}
\setsymbol{HFI:Ndetectors:545GHz}{3}
\setsymbol{HFI:Ndetectors:857GHz}{4}

\setsymbol{HFI:FWHM:Maps:100GHz:units}{9\parcm88}
\setsymbol{HFI:FWHM:Maps:143GHz:units}{7\parcm18}
\setsymbol{HFI:FWHM:Maps:217GHz:units}{4\parcm87}
\setsymbol{HFI:FWHM:Maps:353GHz:units}{4\parcm65}
\setsymbol{HFI:FWHM:Maps:545GHz:units}{4\parcm72}
\setsymbol{HFI:FWHM:Maps:857GHz:units}{4\parcm39}
\setsymbol{HFI:FWHM:Maps:100GHz}{9.88}
\setsymbol{HFI:FWHM:Maps:143GHz}{7.18}
\setsymbol{HFI:FWHM:Maps:217GHz}{4.87}
\setsymbol{HFI:FWHM:Maps:353GHz}{4.65}
\setsymbol{HFI:FWHM:Maps:545GHz}{4.72}
\setsymbol{HFI:FWHM:Maps:857GHz}{4.39}

\setsymbol{HFI:beam:ellipticity:Maps:100GHz}{1.15}
\setsymbol{HFI:beam:ellipticity:Maps:143GHz}{1.01}
\setsymbol{HFI:beam:ellipticity:Maps:217GHz}{1.06}
\setsymbol{HFI:beam:ellipticity:Maps:353GHz}{1.05}
\setsymbol{HFI:beam:ellipticity:Maps:545GHz}{1.14}
\setsymbol{HFI:beam:ellipticity:Maps:857GHz}{1.19}

\setsymbol{HFI:FWHM:Mars:100GHz:units}{9\parcm37}
\setsymbol{HFI:FWHM:Mars:143GHz:units}{7\parcm04}
\setsymbol{HFI:FWHM:Mars:217GHz:units}{4\parcm68}
\setsymbol{HFI:FWHM:Mars:353GHz:units}{4\parcm43}
\setsymbol{HFI:FWHM:Mars:545GHz:units}{3\parcm80}
\setsymbol{HFI:FWHM:Mars:857GHz:units}{3\parcm67}

\setsymbol{HFI:FWHM:Mars:100GHz}{9.37}
\setsymbol{HFI:FWHM:Mars:143GHz}{7.04}
\setsymbol{HFI:FWHM:Mars:217GHz}{4.68}
\setsymbol{HFI:FWHM:Mars:353GHz}{4.43}
\setsymbol{HFI:FWHM:Mars:545GHz}{3.80}
\setsymbol{HFI:FWHM:Mars:857GHz}{3.67}

\setsymbol{HFI:beam:ellipticity:Mars:100GHz}{1.18}
\setsymbol{HFI:beam:ellipticity:Mars:143GHz}{1.03}
\setsymbol{HFI:beam:ellipticity:Mars:217GHz}{1.14}
\setsymbol{HFI:beam:ellipticity:Mars:353GHz}{1.09}
\setsymbol{HFI:beam:ellipticity:Mars:545GHz}{1.25}
\setsymbol{HFI:beam:ellipticity:Mars:857GHz}{1.03}

\setsymbol{HFI:CMB:relative:calibration:100GHz}{$\lsim 1\%$}
\setsymbol{HFI:CMB:relative:calibration:143GHz}{$\lsim 1\%$}
\setsymbol{HFI:CMB:relative:calibration:217GHz}{$\lsim 1\%$}
\setsymbol{HFI:CMB:relative:calibration:353GHz}{$\lsim 1\%$}
\setsymbol{HFI:CMB:relative:calibration:545GHz}{}
\setsymbol{HFI:CMB:relative:calibration:857GHz}{}

\setsymbol{HFI:CMB:absolute:calibration:100GHz}{$\lsim 2\%$}
\setsymbol{HFI:CMB:absolute:calibration:143GHz}{$\lsim 2\%$}
\setsymbol{HFI:CMB:absolute:calibration:217GHz}{$\lsim 2\%$}
\setsymbol{HFI:CMB:absolute:calibration:353GHz}{$\lsim 2\%$}
\setsymbol{HFI:CMB:absolute:calibration:545GHz}{}
\setsymbol{HFI:CMB:absolute:calibration:857GHz}{}

\setsymbol{HFI:FIRAS:gain:calibration:accuracy:statistical:100GHz}{}
\setsymbol{HFI:FIRAS:gain:calibration:accuracy:statistical:143GHz}{}
\setsymbol{HFI:FIRAS:gain:calibration:accuracy:statistical:217GHz}{}
\setsymbol{HFI:FIRAS:gain:calibration:accuracy:statistical:353GHz}{2.5\%}
\setsymbol{HFI:FIRAS:gain:calibration:accuracy:statistical:545GHz}{1\%}
\setsymbol{HFI:FIRAS:gain:calibration:accuracy:statistical:857GHz}{0.5\%}

\setsymbol{HFI:FIRAS:gain:calibration:accuracy:systematic:100GHz}{}
\setsymbol{HFI:FIRAS:gain:calibration:accuracy:systematic:143GHz}{}
\setsymbol{HFI:FIRAS:gain:calibration:accuracy:systematic:217GHz}{}
\setsymbol{HFI:FIRAS:gain:calibration:accuracy:systematic:353GHz}{}
\setsymbol{HFI:FIRAS:gain:calibration:accuracy:systematic:545GHz}{7\%}
\setsymbol{HFI:FIRAS:gain:calibration:accuracy:systematic:857GHz}{7\%}

\setsymbol{HFI:FIRAS:zero:point:accuracy:100GHz:units}{0.8\MJysr}
\setsymbol{HFI:FIRAS:zero:point:accuracy:143GHz:units}{}
\setsymbol{HFI:FIRAS:zero:point:accuracy:217GHz:units}{}
\setsymbol{HFI:FIRAS:zero:point:accuracy:353GHz:units}{1.4\MJysr}
\setsymbol{HFI:FIRAS:zero:point:accuracy:545GHz:units}{2.2\MJysr}
\setsymbol{HFI:FIRAS:zero:point:accuracy:857GHz:units}{1.7\MJysr}

\setsymbol{HFI:FIRAS:zero:point:accuracy:100GHz}{0.8}
\setsymbol{HFI:FIRAS:zero:point:accuracy:143GHz}{}
\setsymbol{HFI:FIRAS:zero:point:accuracy:217GHz}{}
\setsymbol{HFI:FIRAS:zero:point:accuracy:353GHz}{1.4}
\setsymbol{HFI:FIRAS:zero:point:accuracy:545GHz}{2.2}
\setsymbol{HFI:FIRAS:zero:point:accuracy:857GHz}{1.7}

\setsymbol{HFI:unit:conversion:100GHz:units}{0.2415\MJysrmK}
\setsymbol{HFI:unit:conversion:143GHz:units}{0.3694\MJysrmK}
\setsymbol{HFI:unit:conversion:217GHz:units}{0.4811\MJysrmK}
\setsymbol{HFI:unit:conversion:353GHz:units}{0.2883\MJysrmK}
\setsymbol{HFI:unit:conversion:545GHz:units}{0.05826\MJysrmK}
\setsymbol{HFI:unit:conversion:857GHz:units}{0.002238\MJysrmK}

\setsymbol{HFI:unit:conversion:100GHz}{0.2415}
\setsymbol{HFI:unit:conversion:143GHz}{0.3694}
\setsymbol{HFI:unit:conversion:217GHz}{0.4811}
\setsymbol{HFI:unit:conversion:353GHz}{0.2883}
\setsymbol{HFI:unit:conversion:545GHz}{0.05826}
\setsymbol{HFI:unit:conversion:857GHz}{0.002238}

\setsymbol{HFI:colour:correction:alpha=-2:V1.01:100GHz}{0.9893}
\setsymbol{HFI:colour:correction:alpha=-2:V1.01:143GHz}{0.9759}
\setsymbol{HFI:colour:correction:alpha=-2:V1.01:217GHz}{1.0007}
\setsymbol{HFI:colour:correction:alpha=-2:V1.01:353GHz}{1.0028}
\setsymbol{HFI:colour:correction:alpha=-2:V1.01:545GHz}{1.0019}
\setsymbol{HFI:colour:correction:alpha=-2:V1.01:857GHz}{0.9889}

\setsymbol{HFI:colour:correction:alpha=0:V1.01:100GHz}{1.0008}
\setsymbol{HFI:colour:correction:alpha=0:V1.01:143GHz}{1.0148}
\setsymbol{HFI:colour:correction:alpha=0:V1.01:217GHz}{0.9909}
\setsymbol{HFI:colour:correction:alpha=0:V1.01:353GHz}{0.9888}
\setsymbol{HFI:colour:correction:alpha=0:V1.01:545GHz}{0.9878}
\setsymbol{HFI:colour:correction:alpha=0:V1.01:857GHz}{1.0014}

\providecommand{\sorthelp}[1]{}

\begin{document}

\title{\Planck\ Intermediate Results. XXXVI. Optical identification and
  redshifts of \Planck\ SZ sources with telescopes at the Canary Islands
  Observatories}

\titlerunning{\Planck\ SZ source catalogue: Optical identification using Canary
  Islands Observatories}

\author{\small
Planck Collaboration: P.~A.~R.~Ade\inst{86}
\and
N.~Aghanim\inst{56}
\and
M.~Arnaud\inst{71}
\and
M.~Ashdown\inst{67, 6}
\and
J.~Aumont\inst{56}
\and
C.~Baccigalupi\inst{85}
\and
A.~J.~Banday\inst{95, 11}
\and
R.~B.~Barreiro\inst{62}
\and
R.~Barrena\inst{61, 19}
\and
N.~Bartolo\inst{31, 63}
% ...
}
\institute{\small
APC, AstroParticule et Cosmologie, Universit\'{e} Paris Diderot, CNRS/IN2P3, CEA/lrfu, Observatoire de Paris, Sorbonne Paris Cit\'{e}, 10, rue Alice Domon et L\'{e}onie Duquet, 75205 Paris Cedex 13, France\goodbreak
\and
Aalto University Mets\"{a}hovi Radio Observatory, P.O. Box 13000, FI-00076 AALTO, Finland\goodbreak
\and
Academy of Sciences of Tatarstan, Bauman Str., 20, Kazan, 420111, Republic of Tatarstan, Russia\goodbreak
\and
African Institute for Mathematical Sciences, 6-8 Melrose Road, Muizenberg, Cape Town, South Africa\goodbreak
\and
Agenzia Spaziale Italiana Science Data Center, Via del Politecnico snc, 00133, Roma, Italy\goodbreak
\and
Astrophysics Group, Cavendish Laboratory, University of Cambridge, J J Thomson Avenue, Cambridge CB3 0HE, U.K.\goodbreak
\and
Astrophysics \& Cosmology Research Unit, School of Mathematics, Statistics \& Computer Science, University of KwaZulu-Natal, Westville Campus, Private Bag X54001, Durban 4000, South Africa\goodbreak
\and
Atacama Large Millimeter/submillimeter Array, ALMA Santiago Central Offices, Alonso de Cordova 3107, Vitacura, Casilla 763 0355, Santiago, Chile\goodbreak
\and
CGEE, SCS Qd 9, Lote C, Torre C, 4$^{\circ}$ andar, Ed. Parque Cidade Corporate, CEP 70308-200, Bras\'{i}lia, DF,Ê Brazil\goodbreak
\and
CITA, University of Toronto, 60 St. George St., Toronto, ON M5S 3H8, Canada\goodbreak
\and
CNRS, IRAP, 9 Av. colonel Roche, BP 44346, F-31028 Toulouse cedex 4, France\goodbreak
\and
California Institute of Technology, Pasadena, California, U.S.A.\goodbreak
\and
Centro de Estudios de F\'{i}sica del Cosmos de Arag\'{o}n (CEFCA), Plaza San Juan, 1, planta 2, E-44001, Teruel, Spain\goodbreak
\and
Computational Cosmology Center, Lawrence Berkeley National Laboratory, Berkeley, California, U.S.A.\goodbreak
\and
Consejo Superior de Investigaciones Cient\'{\i}ficas (CSIC), Madrid, Spain\goodbreak
\and
DSM/Irfu/SPP, CEA-Saclay, F-91191 Gif-sur-Yvette Cedex, France\goodbreak
\and
DTU Space, National Space Institute, Technical University of Denmark, Elektrovej 327, DK-2800 Kgs. Lyngby, Denmark\goodbreak
\and
D\'{e}partement de Physique Th\'{e}orique, Universit\'{e} de Gen\`{e}ve, 24, Quai E. Ansermet,1211 Gen\`{e}ve 4, Switzerland\goodbreak
\and
Departamento de Astrof\'{i}sica, Universidad de La Laguna (ULL), E-38206 La Laguna, Tenerife, Spain\goodbreak
\and
Departamento de F\'{\i}sica, Universidad de Oviedo, Avda. Calvo Sotelo s/n, Oviedo, Spain\goodbreak
\and
Department of Astronomy and Geodesy, Kazan Federal University,  Kremlevskaya Str., 18, Kazan, 420008, Russia\goodbreak
\and
Department of Astrophysics/IMAPP, Radboud University Nijmegen, P.O. Box 9010, 6500 GL Nijmegen, The Netherlands\goodbreak
\and
Department of Physics \& Astronomy, University of British Columbia, 6224 Agricultural Road, Vancouver, British Columbia, Canada\goodbreak
\and
Department of Physics and Astronomy, Dana and David Dornsife College of Letter, Arts and Sciences, University of Southern California, Los Angeles, CA 90089, U.S.A.\goodbreak
\and
Department of Physics and Astronomy, University College London, London WC1E 6BT, U.K.\goodbreak
\and
Department of Physics, Florida State University, Keen Physics Building, 77 Chieftan Way, Tallahassee, Florida, U.S.A.\goodbreak
\and
Department of Physics, Gustaf H\"{a}llstr\"{o}min katu 2a, University of Helsinki, Helsinki, Finland\goodbreak
\and
Department of Physics, Princeton University, Princeton, New Jersey, U.S.A.\goodbreak
\and
Department of Physics, University of California, Santa Barbara, California, U.S.A.\goodbreak
\and
Department of Physics, University of Illinois at Urbana-Champaign, 1110 West Green Street, Urbana, Illinois, U.S.A.\goodbreak
\and
Dipartimento di Fisica e Astronomia G. Galilei, Universit\`{a} degli Studi di Padova, via Marzolo 8, 35131 Padova, Italy\goodbreak
\and
Dipartimento di Fisica e Scienze della Terra, Universit\`{a} di Ferrara, Via Saragat 1, 44122 Ferrara, Italy\goodbreak
\and
Dipartimento di Fisica, Universit\`{a} La Sapienza, P. le A. Moro 2, Roma, Italy\goodbreak
\and
Dipartimento di Fisica, Universit\`{a} degli Studi di Milano, Via Celoria, 16, Milano, Italy\goodbreak
\and
Dipartimento di Matematica, Universit\`{a} di Roma Tor Vergata, Via della Ricerca Scientifica, 1, Roma, Italy\goodbreak
\and
Discovery Center, Niels Bohr Institute, Blegdamsvej 17, Copenhagen, Denmark\goodbreak
\and
Discovery Center, Niels Bohr Institute, Copenhagen University, Blegdamsvej 17, Copenhagen, Denmark\goodbreak
\and
European Southern Observatory, ESO Vitacura, Alonso de Cordova 3107, Vitacura, Casilla 19001, Santiago, Chile\goodbreak
\and
European Space Agency, ESAC, Planck Science Office, Camino bajo del Castillo, s/n, Urbanizaci\'{o}n Villafranca del Castillo, Villanueva de la Ca\~{n}ada, Madrid, Spain\goodbreak
\and
Facolt\`{a} di Ingegneria, Universit\`{a} degli Studi e-Campus, Via Isimbardi 10, Novedrate (CO), 22060, Italy\goodbreak
\and
Finnish Centre for Astronomy with ESO (FINCA), University of Turku, V\"{a}is\"{a}l\"{a}ntie 20, FIN-21500, Piikki\"{o}, Finland\goodbreak
\and
Gran Sasso Science Institute, INFN, viale F. Crispi 7, 67100 L'Aquila, Italy\goodbreak
\and
HGSFP and University of Heidelberg, Theoretical Physics Department, Philosophenweg 16, 69120, Heidelberg, Germany\goodbreak
\and
Helsinki Institute of Physics, Gustaf H\"{a}llstr\"{o}min katu 2, University of Helsinki, Helsinki, Finland\goodbreak
\and
INAF - Osservatorio Astronomico di Padova, Vicolo dell'Osservatorio 5, Padova, Italy\goodbreak
\and
INAF - Osservatorio Astronomico di Roma, via di Frascati 33, Monte Porzio Catone, Italy\goodbreak
\and
INAF - Osservatorio Astronomico di Trieste, Via G.B. Tiepolo 11, Trieste, Italy\goodbreak
\and
INAF/IASF Bologna, Via Gobetti 101, Bologna, Italy\goodbreak
\and
INAF/IASF Milano, Via E. Bassini 15, Milano, Italy\goodbreak
\and
INFN, Sezione di Bologna, Via Irnerio 46, I-40126, Bologna, Italy\goodbreak
\and
INFN, Sezione di Roma 1, Universit\`{a} di Roma Sapienza, Piazzale Aldo Moro 2, 00185, Roma, Italy\goodbreak
\and
INFN, Sezione di Roma 2, Universit\`{a} di Roma Tor Vergata, Via della Ricerca Scientifica, 1, Roma, Italy\goodbreak
\and
Imperial College London, Astrophysics group, Blackett Laboratory, Prince Consort Road, London, SW7 2AZ, U.K.\goodbreak
\and
Infrared Processing and Analysis Center, California Institute of Technology, Pasadena, CA 91125, U.S.A.\goodbreak
\and
Institut Universitaire de France, 103, bd Saint-Michel, 75005, Paris, France\goodbreak
\and
Institut d'Astrophysique Spatiale, CNRS (UMR8617) Universit\'{e} Paris-Sud 11, B\^{a}timent 121, Orsay, France\goodbreak
\and
Institut d'Astrophysique de Paris, CNRS (UMR7095), 98 bis Boulevard Arago, F-75014, Paris, France\goodbreak
\and
Institute of Astronomy, University of Cambridge, Madingley Road, Cambridge CB3 0HA, U.K.\goodbreak
\and
Institute of Theoretical Astrophysics, University of Oslo, Blindern, Oslo, Norway\goodbreak
\and
Instituto Nacional de Astrof\'{\i}sica, \'{O}ptica y Electr\'{o}nica (INAOE), Apartado Postal 51 y 216, 72000 Puebla, M\'{e}xico\goodbreak
\and
Instituto de Astrof\'{\i}sica de Canarias, C/V\'{\i}a L\'{a}ctea s/n, La Laguna, Tenerife, Spain\goodbreak
\and
Instituto de F\'{\i}sica de Cantabria (CSIC-Universidad de Cantabria), Avda. de los Castros s/n, Santander, Spain\goodbreak
\and
Istituto Nazionale di Fisica Nucleare, Sezione di Padova, via Marzolo 8, I-35131 Padova, Italy\goodbreak
\and
Jet Propulsion Laboratory, California Institute of Technology, 4800 Oak Grove Drive, Pasadena, California, U.S.A.\goodbreak
\and
Jodrell Bank Centre for Astrophysics, Alan Turing Building, School of Physics and Astronomy, The University of Manchester, Oxford Road, Manchester, M13 9PL, U.K.\goodbreak
\and
Kavli Institute for Cosmological Physics, University of Chicago, Chicago, IL 60637, USA\goodbreak
\and
Kavli Institute for Cosmology Cambridge, Madingley Road, Cambridge, CB3 0HA, U.K.\goodbreak
\and
Kazan Federal University, 18 Kremlyovskaya St., Kazan, 420008, Russia\goodbreak
\and
LAL, Universit\'{e} Paris-Sud, CNRS/IN2P3, Orsay, France\goodbreak
\and
LERMA, CNRS, Observatoire de Paris, 61 Avenue de l'Observatoire, Paris, France\goodbreak
\and
Laboratoire AIM, IRFU/Service d'Astrophysique - CEA/DSM - CNRS - Universit\'{e} Paris Diderot, B\^{a}t. 709, CEA-Saclay, F-91191 Gif-sur-Yvette Cedex, France\goodbreak
\and
Laboratoire Traitement et Communication de l'Information, CNRS (UMR 5141) and T\'{e}l\'{e}com ParisTech, 46 rue Barrault F-75634 Paris Cedex 13, France\goodbreak
\and
Laboratoire de Physique Subatomique et Cosmologie, Universit\'{e} Grenoble-Alpes, CNRS/IN2P3, 53, rue des Martyrs, 38026 Grenoble Cedex, France\goodbreak
\and
Laboratoire de Physique Th\'{e}orique, Universit\'{e} Paris-Sud 11 \& CNRS, B\^{a}timent 210, 91405 Orsay, France\goodbreak
\and
Lawrence Berkeley National Laboratory, Berkeley, California, U.S.A.\goodbreak
\and
Lebedev Physical Institute of the Russian Academy of Sciences, Astro Space Centre, 84/32 Profsoyuznaya st., Moscow, GSP-7, 117997, Russia\goodbreak
\and
Max-Planck-Institut f\"{u}r Astrophysik, Karl-Schwarzschild-Str. 1, 85741 Garching, Germany\goodbreak
\and
Max-Planck-Institut f\"{u}r Extraterrestrische Physik, Giessenbachstra{\ss}e, 85748 Garching, Germany\goodbreak
\and
Moscow Institute of Physics and Technology, Dolgoprudny, Institutsky per., 9, 141700, Russia\goodbreak
\and
National University of Ireland, Department of Experimental Physics, Maynooth, Co. Kildare, Ireland\goodbreak
\and
Nicolaus Copernicus Astronomical Center, Bartycka 18, 00-716 Warsaw, Poland\goodbreak
\and
Niels Bohr Institute, Blegdamsvej 17, Copenhagen, Denmark\goodbreak
\and
Niels Bohr Institute, Copenhagen University, Blegdamsvej 17, Copenhagen, Denmark\goodbreak
\and
Optical Science Laboratory, University College London, Gower Street, London, U.K.\goodbreak
\and
SISSA, Astrophysics Sector, via Bonomea 265, 34136, Trieste, Italy\goodbreak
\and
School of Physics and Astronomy, Cardiff University, Queens Buildings, The Parade, Cardiff, CF24 3AA, U.K.\goodbreak
\and
Sorbonne Universit\'{e}-UPMC, UMR7095, Institut d'Astrophysique de Paris, 98 bis Boulevard Arago, F-75014, Paris, France\goodbreak
\and
Space Research Institute (IKI), Russian Academy of Sciences, Profsoyuznaya Str, 84/32, Moscow, 117997, Russia\goodbreak
\and
Space Sciences Laboratory, University of California, Berkeley, California, U.S.A.\goodbreak
\and
Special Astrophysical Observatory, Russian Academy of Sciences, Nizhnij Arkhyz, Zelenchukskiy region, Karachai-Cherkessian Republic, 369167, Russia\goodbreak
\and
Sub-Department of Astrophysics, University of Oxford, Keble Road, Oxford OX1 3RH, U.K.\goodbreak
\and
T\"{U}B\.{I}TAK National Observatory, Akdeniz University Campus, 07058, Antalya, Turkey\goodbreak
\and
UPMC Univ Paris 06, UMR7095, 98 bis Boulevard Arago, F-75014, Paris, France\goodbreak
\and
Universidad Andr\'{e}s Bello, Dpto. de Ciencias F\'{i}sicas, Facultad de Ciencias Exactas, 8370134 Santiago de Chile, Chile\goodbreak
\and
Universit\'{e} de Toulouse, UPS-OMP, IRAP, F-31028 Toulouse cedex 4, France\goodbreak
% ...
}

\authorrunning{Planck Collaboration}


\abstract{ 

We present the results of approximately three years of observations of
\Planck\ Sunyaev-Zeldovich (SZ) sources with telescopes at the Canary Islands
observatories as part of the general optical follow-up programme undertaken by
the \Planck\ collaboration.
In total, 78 SZ sources are discussed. Deep-imaging observations were obtained
for most of these sources; spectroscopic observations in either in long-slit or
multi-object modes were obtained for many. We effectively
used $37.5 $ clear nights.
We found optical counterparts for 73 of the 78 candidates. This sample includes
53 spectroscopic redshift determinations, 20 of them obtained with a
multi-object spectroscopic mode. The sample contains new redshifts for 27
\Planck\ clusters that were not included in the first \Planck\ SZ source
catalogue (PSZ1).

}


\maketitle

\section{Introduction}

The Sunyaev-Zeldovich (SZ) effect \citep{sz1972} is a spectral distortion of the
cosmic microwave background (CMB) that is generated by inverse Compton scattering of CMB
photons off hot electrons. It can be used as a powerful cosmological probe,
complementary to the information encoded in the CMB angular power spectrum
\citep[e.g.][]{birkinshaw1999, carlstrom2002, planck2013-p15, planck2013-p05b,
  planck2013-p05a}.
The usual application of the SZ effect is the detection of galaxy clusters,
which in the standard paradigm are manifested in the most massive halos that
emerge in the cosmic web of the large-scale structure
\citep[e.g.][]{springel05}. These halos are multi-component systems, consisting
of dark matter and baryons in several phases \citep[e.g.][]{allen11}. Galaxy
clusters are excellent cosmological probes, capable of testing cosmology in a
variety of ways, and placing constraints on cosmological parameters such as the
normalization of the matter spectrum, dark matter and dark energy densities,
neutrino masses, and the equation of state of the dark energy
\citep{vikhlinin09,Henry2009, mantz10a, planck2013-p15, mantz2015,
  planck2014-a30}. Of all galaxy clusters, the most massive ones are the most
sensitive to the cosmology. However, in the $\Lambda$CDM model, these most
massive galaxy clusters are predicted to be very rare, and their abundance
rapidly decreases with redshift \citep[e.g.][]{springel05}.
Observationally, therefore, they are best found by surveying the large volumes
accessible with all-sky surveys.

In the past, such surveys have been conducted using either X-ray or optical
observations. Examples of the former are the ROSAT All-Sky Survey (RASS) and the
catalogues drawn from RASS (e.g. REFLEX \citep{reflex}, NORAS \citep{noras}, or
MACS \citep{macs}), which reach a maximum redshift of $z\sim 0.5$ with a few
objects beyond. Examples of the latter include catalogues based on SDSS data
\citep{koester2007b,wen2009,hao2010,szabo2010,WHL2012}, which extend up to about
the same redshift threshold. In the past few years, millimeter surveys have
released the first sets of galaxy clusters discovered by means of the thermal SZ
effect \citep[e.g.][]{2012ApJ...761...22S, 2013ApJ...763..127R,
  2013JCAP...07..008H, 2013ApJ...772...25S}.

The \Planck\footnote{\Planck\ (\url{http://www.esa.int/Planck}) is a project of
  the European Space Agency (ESA) with instruments provided by two scientific
  consortia funded by ESA member states and led by Principal Investigators from
  France and Italy, telescope reflectors provided through a collaboration
  between ESA and a scientific consortium led and funded by Denmark, and
  additional contributions from NASA (USA). } satellite \citep{planck2013-p01}
provided for the first time the possibility of detecting galaxy clusters through
the SZ effect in a full sky survey
\citep{planck2011-5.1a,planck2013-p05a,planck2014-a36}. However, as the SZ
effect provides no redshift, dedicated follow-up programmes are required to make
the resulting catalogues scientifically useful. Since 2010, the
\Planck\ collaboration has undertaken an extensive follow-up programme to
confirm SZ cluster candidates, first from intermediate versions of the
\Planck\ SZ catalogue \citep{planck2011-5.1b,planck2012-I,planck2012-IV}, and
later with the first public SZ catalogue, named PSZ1
\citep{planck2013-p05a,planck2013-p05a-addendum}. The strategy of the follow-up
programme is detailed in that PSZ1 paper.

In this paper, we describe the observations carried out with telescopes at the
Canary Islands Observatories (or the European Northern Observatory,
ENO\footnote{\url{http://www.iac.es/eno.php?lang=en}}), which were used for the
PSZ1 catalogue \citep{planck2013-p05a,planck2013-p05a-addendum}. This is the
second companion paper to the PSZ1 that is based on optical follow-up
observations. The first was based on observations with the Russian-Turkish
1.5\,m telescope (RTT; \citealt{planck2014-XXVI}). Clusters confirmed in that
work and here are tabulated in \cite{planck2014-a36}.

The paper is organized as follows. Sections~\ref{sec:sample},
\ref{sec:followup}, and \ref{sec:methods} describe the \Planck\ catalogue and
cluster observing programme, the telescopes used in this paper, and the
methods we used to analyse our data and identify clusters. Section~\ref{sec:results} describes the observations and main
results. In Sect.~\ref{sec:conclusions} we conclude.

\section{Planck PSZ1 cluster sample}
\label{sec:sample}

The first \Planck\ catalogue of SZ sources (PSZ1;
\citealt{planck2013-p05a,planck2013-p05a-addendum}) comprises 1227 objects
detected by means of the SZ effect in the \Planck\ all-sky maps of the first
15.5~months of observations. The method used to construct the PSZ1
catalogue is described in detail in \citet{planck2013-p05a}. In brief, cluster
candidates are blindly detected using three different detection methods (MMF1,
MMF3, and PwS), which were also used to construct the ESZ sample
\citep{planck2011-5.1a}. The PSZ1 catalogue contains all objects found by at
least one of the methods with a signal-to-noise ratio (S/N) of 4.5 or greater.

It is important to recall here that the mean beam size of the HFI \Planck\ maps
extends from $9\parcm6$ at 100\,GHz to $4\parcm5$ at 857\,GHz \citep[see
  e.g.][]{planck2013-p01}. However, the expected positional error for the whole
PSZ1 sample is $2\arcm$ \citep{planck2013-p05a}.

A detailed validation process \citep{planck2013-p05a} included the search for
possible counterparts in the ROSAT All Sky Survey
\citep[RASS,][]{rassbr,rassfaint}, the Sloan Digital Sky Survey \citep[SDSS,
  DR8,][]{2011ApJS..193...29A}, the {\it WISE} all-sky survey \citep{wise}, and
DSS\footnote{DSS: \url{http://stdatu.stsci.edu/dss}} images, as well as
cross-correlations with optical, X-ray, and other SZ catalogues. Candidates
without confirmation after these steps were sent to observing facilities for
follow-up observations. This article is one of the companion papers to
\citet{planck2013-p05a} and the PSZ1 catalogue.

We finally note that the \Planck\ collaboration recently released its second
catalogue of SZ sources (PSZ2; \citealt{planck2014-a36}). PSZ2 includes 22
confirmations that are common sources with the PSZ1 and are discussed in this paper.

\section{Our follow-up observations}
\label{sec:followup}

Table~\ref{tab:telescopes} lists the telescopes and instruments at the Canary
Islands Observatories that have been used for this paper: a) the Gran Telescopio
Canarias (GTC), installed in the Spanish Observatorio del Roque de los Muchachos
(ORM) of the Instituto de Astrof\'{\i}sica de Canarias (IAC), in the island of
La Palma; b) the Isaac Newton Telescope and the William Herschel Telescope
operated on the island of La Palma by the ISAAC Newton Group of Telescopes in
the Spanish ORM of the IAC; c) the Italian Telescopio Nazionale Galileo (TNG)
operated on the island of La Palma by the Fundaci\'on Galileo Galilei of the
INAF (Istituto Nazionale di Astrofisica) at the Spanish ORM of the IAC; d) the
Nordic Optical Telescope, jointly operated on the island of La Palma by Denmark,
Finland, Iceland, Norway, and Sweden, in the Spanish ORM of the IAC; and e) the
IAC80 telescope operated on the island of Tenerife by the IAC in the Spanish
Observatorio del Teide.
The observed targets were extracted from different internal versions of the
\Planck\ cluster catalogue before publication of the PSZ1. The main
criterion for observing a certain target in a given right ascension range was
the $S/N$ of the \Planck\ SZ detection. We also note that the lowest declination
accessible from the Canary Islands at a reasonable airmass is $-20\deg$. Targets
with lower declinations are sent to other facilities in the Southern Hemisphere.

\begin{table*}[h!]
\begingroup
\newdimen\tblskip \tblskip=5pt
\caption{List of the telescopes and instruments used for the confirmation of the
  \Planck\ newly-discovered clusters at the Canary Islands Observatories. For
  spectroscopy, the column 7 shows the resolution. Last column indicates
    the total number of clear nights (equivalent to 9~hours) observed in each
    telescope for this work.}
\label{tab:telescopes}
\nointerlineskip 
\vskip -2mm 
\footnotesize
\setbox\tablebox=\vbox{
\newdimen\digitwidth
\setbox0=\hbox{\rm 0}
\digitwidth=\wd0
\catcode`*=\active
\def*{\kern\digitwidth}
\newdimen\signwidth
\setbox0=\hbox{+}
\signwidth=\wd0
\catcode`!=\active
\def!{\kern\signwidth}
\halign{\hbox to 1.3in{#\leaderfil}\tabskip=2em&
        \hfil#\hfil& 
        \hfil#\hfil&
        \hfil#\hfil&
        \hfil#\hfil&
        \hfil#\hfil&
        \hfil#\hfil&
        \hfil#\hfil\tabskip=0pt\cr
\noalign{\doubleline}
\omit\hfil Site\hfil& Telescope& Aperture [m]& Instrument& FoV& Pixel Scale
           [$\arcs$]& Resolution & Nights \cr
\noalign{\vskip 3pt\hrule\vskip 5pt}
OT, Tenerife& IAC80& *0.82& CAMELOT&  $10\parcm4 \times 10\parcm4$& 0.304& $\dots$&12**\cr
ORM, La Palma&  NOT& *2.56&  ALFOSC&        $7\arcm \times 7\arcm$& 0.188& $R=550$&*0.5\cr
ORM, La Palma&  INT& *2.5*& WFC/IDS&      $34\arcm \times 34\arcm$& 0.33*& $R=400$&15**\cr
ORM, La Palma&  TNG& *3.5*& DOLORES&    $8\parcm6 \times 8\parcm6$& 0.252& $R=600$&*5**\cr
ORM, La Palma&  WHT& *4.2*&    ACAM&        $8\arcm \times 8\arcm$& 0.253& $R=400$&*2**\cr
ORM, La Palma&  GTC& 10.4*&  OSIRIS&   $7\parcm8 \times 7\parcm 8$& 0.254& $R=500$&*3**\cr
\noalign{\vskip 5pt\hrule\vskip 3pt}}}
\endPlancktable
\endgroup
\end{table*}

For each cluster candidate, we performed an initial pre-screening of possible
counterparts using the Digitized Sky Survey
(DSS)\footnote{\url{http://archive.stsci.edu/dss}} and the Sloan Digital Sky
Survey (SDSS)\footnote{\url{http://skyserver.sdss.org}} , if available.
If a cluster counterpart was identified after this process, new imaging
observations were not required. If SDSS spectroscopic observations were available
for several cluster members, we can provide the spectroscopic redshift of
the cluster. Otherwise, we included the cluster in our list of spectroscopic
targets.

After a certain target was selected for imaging observations, it was examined in
two steps with our follow-up strategy. First, we carried out photometric
observations in at least three broadband Sloan filters
($g^\prime$,$r^\prime$,$i^\prime$). In some particular cases, we also used a
redder filter (either Sloan $z^\prime$ or Gunn $Z$). Given the expected redshift
distribution of the \Planck\ SZ-detected clusters \citep[see
  e.g.][]{planck2013-p05a}, this filter set is sufficient to trace the
4000\,$\AA$ break up to redshifts $z=0.8$, even in cases where no
$z^\prime$-band data are available. Based on $g^\prime r^\prime i^\prime$-colour
combinations, it is possible to identify the cluster candidates and to measure
photometric redshifts \citep[see e.g.][]{lopes07}.
In a second step, these results were later consolidated using spectroscopic
observations, either using long-slit or multi-object spectroscopy.

Most of the targets were included in the PSZ1 catalogue and are
therefore identified with their corresponding name in this catalogue.  Nevertheless, there
are 11 confirmed targets that were not included in PSZ1, either because
their $S/N$ value is lower than 4.5 or because they were excluded by the
detection mask used in the final catalogue.

\subsection{Imaging observations}

Imaging observations were taken in multiple observing runs between June 2010 and
December 2012, during the \Planck\ proprietary period, mainly with three
instruments: the Wide-Field Camera (WFC) on the 2.5\,m Isaac Newton Telescope
(INT); the auxiliary-port camera (ACAM) at the 4.2\,m William-Herschel
Telescope; and CAMELOT, the optical camera at the 0.82\,m telescope (IAC80).
The observing time was obtained as part of the proposals for the Spanish Time
Allocation Committee (CAT) time (semesters 2010A to 2012B), and an International
Time Programme (ITP12-2), accepted by the International Scientific Committee of
the Roque de los Muchachos (ORM, La Palma) and Teide (OT, Tenerife)
observatories.

All fields were imaged with the $g^\prime$, $r^\prime$ and $i^\prime$ Sloan
filters; for a majority of fields, either Sloan $z^\prime$ or Gunn $Z$ images
are also available. As we discuss below, our final analysis for photometric
redshifts does not make use of the $z^\prime$ band images.

The WFC at the INT is a mosaic camera of four $2\,{\rm k}\times 4\,{\rm k}$ CCDs
with a spatial scale of $0\parcs333/$pixel, resulting in a field of view of
$34\arcm \times 34\arcm$. The detector of ACAM at the WHT covers a circular
field of about $8\parcm3$ (with $\sim 0\parcs25/$pixel), and CAMELOT at the
IAC80 has a $2\,{\rm k}\times 2\,{\rm k}$ chip with $0\parcs304/$pixel spatial
scale, corresponding to a $10\parcm4 \times 10\parcm4$ field. A typical field
size of $10\arcm$ is sufficiently large to cover the expected uncertainty in
positions with respect to the nominal \Planck\ coordinates for each target
\citep{planck2013-p05a}.

The total integration time was split into three separate exposures, with offsets
between exposures of at least $10\arcs$.  This allowed us to efficiently correct for bad
pixels and cosmic rays. The average exposure times were 3000\,s,
1500\,s, and 900\,s per filter for IAC80, INT, and WHT, respectively, yielding
typical completeness magnitudes (derived from the histograms of objects)
in $r^\prime$ band of 20.6, 21.8, and 22.6, and typical limit magnitudes of
21.8, 22.7, and 23.7 for IAC80, INT and WHT, respectively.

\subsection{Spectroscopic observations}

Spectroscopic observations were performed under an International Time Program
(ITP\footnote{ITP: \url{http://www.iac.es/eno.php?op1=5&op2=13&lang=en}}) of the
ORM (reference ITP12-2) and other CAT runs between
2012 and 2013. The redshift information included in this paper was supplemented
with spectroscopic observations of a second ITP program (reference ITP13-08).
The data were mostly obtained with the 3.5\,m TNG and the 10.4\,m GTC
telescopes, while some data were obtained with the 2.5\,m INT and NOT telescopes
in service mode. Multi-object spectroscopy mode (MOS) was used for all the
observations at the TNG telescope and for some of the GTC observations. INT and
NOT observations, as well as most of the GTC observing blocks, were made with
long slits.

The instrument used at the TNG was DOLORES, which has a CCD of $2048\times 2048$
pixels, with pixel size 13.5\,$\mu$m and pixel scale $0\parcs252$. We used the
LR-B grism with a dispersion of 2.75\,$\AA$\,pixel\mo\ operating in the
wavelength range 3000--8430\,$\AA$, together with $1\parcs6$ slits, which led to
a final resolution of $R=600$. The typical exposure time was
$3\times1800$\,s per mask. On average, we placed 30--35 slits per
mask. Wavelength calibration was obtained using Hg-Ne and He-Ne lamps, which
allowed us to obtain an rms error lower than 0.1\,$\AA$\,pixel\mo\ over the
whole wavelength range.

Most of the GTC spectra were obtained using the OSIRIS spectrograph in long-slit
mode. OSIRIS is a double CCD ($2048\times 4096$ pixels) with a pixel size of
15\,$\mu$m and $0\parcs13$ pixel scale. We used the R300B grism, which provides
a dispersion of 5.2\,$\AA$\,pixel\mo\ (using $2\times2$ binning) operating in
the wavelength range 4000--9000\,$\AA$, together with $1\parcs2$ slits. This
set-up produces a spectral resolution of $R \sim 500$. The typical
exposure time was $3\times1000$\,s per slit position, and wavelength calibration
was obtained using Hg, Ne, and Ar lamps, which produced an accuracy of
0.2\,$\AA$\,pixel\mo\ in the whole wavelength range.
For specific cases, we also used the MOS mode with OSIRIS to sample the galaxy
population and velocity field of the clusters. We chose the same instrument
set-up as in long-slit mode, that is, $1\parcs2$ slitlets and the R300B
grism. The MOS mode allows us to set about 35--40 slitlets per mask plus a
minimum of five fiducial holes, which are used to correctly centre the mask in
the field. Masks were designed using GTC pre-imaging for each field. Given that
we previously obtained photometric information (we mainly considered colours for
cluster candidates within the red-sequence and RGB colour composite images
obtained by the combination of $g^\prime$, $r^\prime$, and $i^\prime$-band
images), the success rate was typically 50--60\%, from which we could retrieve
velocity estimates for a minimum of 15--20 cluster members per mask.

The INT and NOT telescopes were also used in service time, with the IDS and
ALFOSC spectrographs in long-slit mode. We used IDS/INT with the R150V grism
with a slit of $1\parcs5$ width, which gave a dispersion of
4.03\,$\AA$\,pixel\mo\ and a spectral resolution of $R = 400$. Similarly, we
used ALFOSC/NOT with grism 4 ($3.05$\, $\AA$\,pixel\mo) together with $1\parcs8$
width (so, $R = 550$). We obtained data exposing $3 \times 1800$\,s in
both telescopes. In both cases we used Ne arcs, obtaining an rms of less than
$0.1$\,$\AA$\,pixel\mo\ across the full spectra

We also used the 4.2\,m WHT telescope with the ACAM spectrograph in long-slit
observing mode, which offers a $2000\times 2000$ CCD with a plate scale of
$0\parcs25$ per pixel. The spectroscopic set-up was configured with a volume
phase holographic disperser (3.3\,$\AA$\,pixel\mo) and a slit width of
$1\arcs$. This set-up produces a spectral coverage between 4000 and 9000\,$\AA$
and a resolution of $R\sim 400$. We usually obtain spectra for the BCG and other
few additional luminous galaxy members by using several slit orientations for
each cluster. On average, we integrate about $2\times 1000$\,s per slit. Our
spectra have $S/N\sim 5$--$30$, which allows us to estimate radial velocities with
an accuracy of about $50$--$80$\,km\,s\mo.


\section{Data reduction, cluster identification, and redshift measurements}
\label{sec:methods}

\subsection{Imaging data}

Optical images were reduced with standard routines running within {\tt
  IRAF}\footnote{{\tt IRAF} (\url{http://iraf.noao.edu/}) is distributed by the
  National Optical Astronomy Observatories, which are operated by the
  Association of Universities for Research in Astronomy, Inc., under cooperative
  agreement with the National Science Foundation.}. The standard reduction
includes flux and astrometric calibrations. For the latter, we used {\tt
  astrometry v0.38}\footnote{http://astrometry.net} \citep{lang2010}. The flux
calibration refers to SDSS and is based either on the observations of
Landolt standard stars or, in the case of fields with SDSS coverage, on the
photometric data of SDSS. In case of non-photometric nights, targets were
calibrated in the following observing run or during service time at the Canary
Islands Observatories.

Sources were detected using {\tt SExtractor} \citep{bertin1996} in single-image
mode. Sources were independently detected in all bands with 1.5$\sigma$
detection thresholds in the filtered maps, which corresponds to a S/N $\sim 3$
in at least ten connected pixels. All catalogues were then merged with a
search radius of $2\arcs$.

The detection limits were estimated by measuring the standard deviation of the
flux in several thousand randomly distributed circular apertures of $3\arcs$
diameter. None of the aperture positions lies within $3\arcs$ of its closest
neighbour or of sources that have a $>3\sigma$ detection.
We note that because of the poor quality of the PSF determination for several
observations, we did not apply any criteria for star/galaxy separation, but
included all sources in our catalogues.

\subsection{Cluster detection and photometric redshifts}
\label{sec:photoz}

Our method to identify the member galaxies in the galaxy cluster candidates and
to derive the photometric redshifts is based on the cluster-red sequence (CRS)
method by \citet{gladders2000}. Colour cuts in $(g^\prime - r^\prime)$ or
$(r^\prime -i^\prime)$ are combined with the spatial distribution of the full
catalogue to search for possible candidates. We have modified this method with
the empirical relations of \citet{lopes07} for the photometric redshifts of
galaxy clusters (see Sect.~4.2 in that paper) to obtain photometric redshift
estimates for galaxy overdensities and to select the probable cluster members
for spectroscopic follow-up observations. To first order, the photometric
redshifts can be estimated with
\begin{eqnarray}
\zphot &\approx 0.361 (g^\prime - r^\prime) - 0.278, \qquad &\hbox{if } (r^\prime - i^\prime) \la 0.75 \\%
\zphot &\approx 0.364 (r^\prime - i^\prime) + 0.182, \qquad &\hbox{if } (r^\prime - i^\prime) \ga 0.75 \rlap{\,.}
\end{eqnarray}
We compared this methodology with the photometric redshifts obtained with
the {\sc BPZ} code \citep{benitez2000} and found that the two methods are fully
consistent for the colour-selected
galaxies in the cluster candidates.

After we applied our method to all candidates, the results were evaluated and
consolidated with a careful inspection of the individual and RGB images and with
the colour-magnitude diagrams. We found that the inspection by eye is
particularly important to detect low-redshift systems, poor systems, fossil
groups, or high-redshift systems, which usually are not easily detected by
automatic algorithms searching for overdensities in colour space.

Our analysis shows that in most of the cases there is an obvious, rich
overdensity of red galaxies in our follow-up imaging. In some cases we found two
separate overdensities; for these, we obtained separate (photometric) redshifts,
assuming that the spatial proximity can be a projection effect along the line of
sight. These cases were investigated in detail using spectroscopic redshifts.

\subsection{Spectroscopic data}

We performed the data reduction of spectroscopic data using standard {\tt IRAF}
procedures. We did not correct for bias or flat fielding because
we found after a careful
check that the uncorrected spectra were less noisy than the corrected ones,
probably because of the low brightness of the sources. The sky was evaluated
locally for each spectrum; we used the median value in the external regions of
each slit. Finally, we looked for possible deviations in the
wavelength-calibrated spectra using the OI telluric line at $5577.3$\,$\AA$.
The mean error associated with this distortion of the instrument is $1$\,$\AA$,
equivalent to about 50\,km\,s$^{-1}$.

The long-slit observations were designed to maximize the number of
redshift estimates, thus placing between two to four galaxies in the slit and retrieving
exposures in two position angles for each cluster candidate. With this scheme,
in some cases we obtained more than five redshifts for a single cluster using
OSIRIS/GTC with only two slit orientations. Furthermore, we always obtained
spectroscopic data of the apparent brightest cluster galaxy (BCG) of each
candidate.

\begin{figure*}[h!]
\centering
\includegraphics[width=\columnwidth]{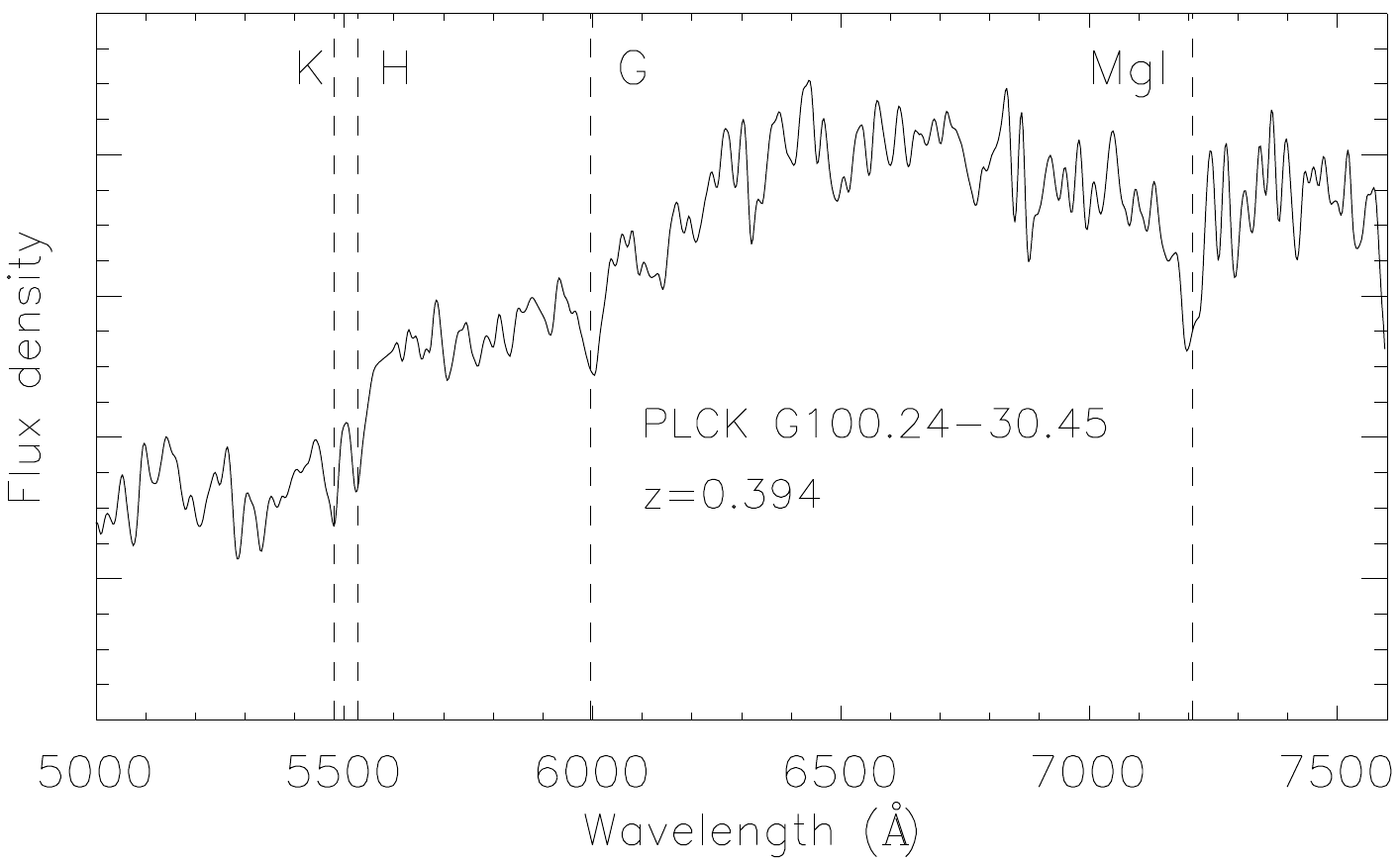}
\includegraphics[width=\columnwidth]{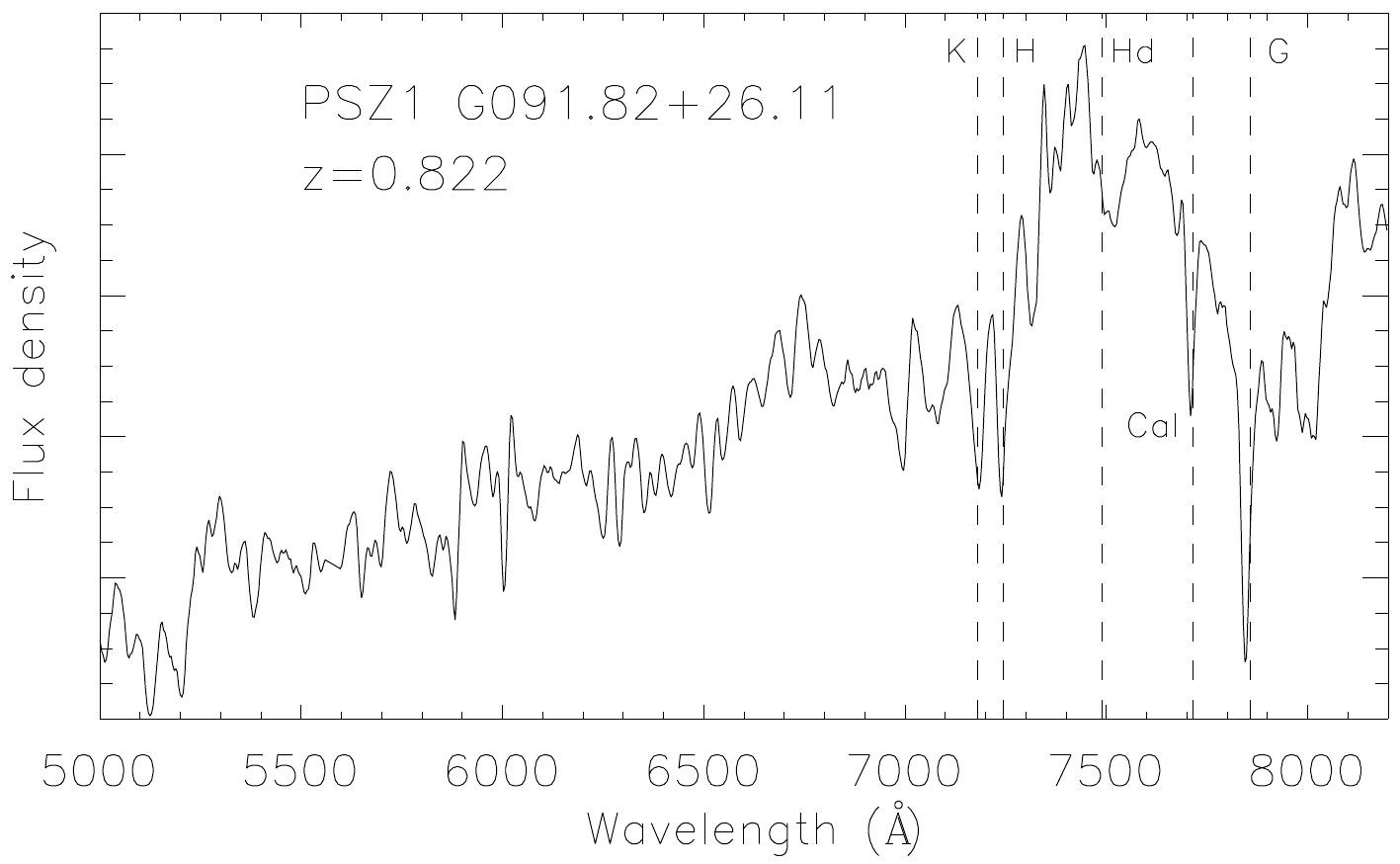}
\caption{\color{black}Two examples of spectra of early-type cluster members
  (previously selected as candidates in the photometric sample) obtained with
  the TNG/DOLORES (left panel) and GTC/OSIRIS (right panel). Dashed lines show
  the location of the main absoption features identified in the spectra. The
  vertical axes are given in arbitrary units. }
\label{fig:xc_spec}
\end{figure*}

After the reduction steps, we obtained spectra with S/N of about
5 (per pixel, and around $6000$\,$\AA$) for galaxies with magnitudes $r^\prime
\sim 20.8, 20.5, 19.7$, and $21.7$ with the WHT, TNG, INT/NOT, and GTC data,
respectively. Figure~\ref{fig:xc_spec} shows two examples of the obtained spectra.

The overall strategy was to use the GTC telescope to observe the most distant
cluster candidates (at photometric redshifts $z_{\rm phot} > 0.4$), while WHT,
INT, and NOT were used for the nearest ones (at $z_{\rm phot} \la 0.4$). We used
TNG/MOS for clusters at intermediate redshift, $0.3\le z_{\rm phot} \le 0.4$.
The MOS mode of OSIRIS at GTC was also used for a few galaxy clusters with
confirmed redshifts above $z=0.5$.

Radial velocities were obtained using the cross-correlation technique
\citep{tonrydavis79} implemented in the {\tt IRAF} task {\tt
  RVSAO}\footnote{RVSA was developed at the Smithsonian Astrophysical
  Observatory Telescope Data Center.}. For each spectrum, the task performs a
cross-correlation with six spectrum templates \citep{kennicutt92}, each one
corresponding to a different type of galaxy: E, S0, Sa, Sb, Sc, and Irr. The
template with the highest $R$-value (a parameter given by XCSAO that measures
the S/N ratio of the cross-correlation peak) was chosen. In addition, we
visually inspected all spectra to verify the velocity determination. In most
cases, the redshift was obtained by using absorption lines. Nevertheless, in
some special cases with low S/N spectra, emission lines were used to determine the
redshift.

The TNG and GTC multi-object observations allowed us to obtain a larger sample
of cluster members for each candidate (typically 15--20 members per
cluster). For these systems we were therefore able to estimate a more accurate mean
velocity, but also other physical magnitudes, like the velocity dispersion of
candidates. In all the other cases, the cluster redshift was assumed to be that
of the BCG (if observed). Otherwise, it was estimated from the mean velocity of
the galaxy members.

Throughout this paper, galaxies are considered cluster members only if they
present radial velocities within $\pm 2000$\,km\,s$^{-1}$ with respect to the
mean velocity of the systems. This velocity range represents about three times
the typical velocity dispersion of a cluster. Hence, this criterium allows us to
select a statistically significant number of cluster galaxies while minimizing
the number of interlopers.

\section{Results}
\label{sec:results}

The list of 67 clusters from the PSZ1 catalogue observed in this paper is given
in Table~\ref{tab:inpsz1}.  An additional set of 11 confirmed clusters not in
the PSZ1 catalogue is given in Table~\ref{tab:nopsz1}.  Fifty-three
spectroscopic redshifts have been measured, 20 obtained with the MOS mode of TNG
or GTC telescopes (i.e. with more than ten confirmed galaxy members).  There are
five SZ sources in the PSZ1 with no confirmed optical counterpart from our
analyses (see discussion below) and 27 new redshifts that were not included in
the original PSZ1 catalogue.
Table~\ref{tab:inpsz1} contains 15 targets that were also observed with the
RTT150 and are discussed in \citet{planck2014-XXVI}. In both analyses, we found
only one difference for a single cluster; this is discussed below.
In addition, two targets (IDs 116 and 316 in Table~\ref{tab:inpsz1}) were also
observed independently in the $r$ and $z$ bands with MegaCam at CFHT
\citep{megacam}. Their photometric redshifts from the $r-z$ red-sequence are
consistent with our spectroscopic redshifts.

\begin{table*}
\begingroup
\caption{Clusters from the PSZ1 catalogue observed for this paper. The first
  column gives the index number of the cluster in the PSZ1 catalogue (starting
  from 0). Column 3 indicates if the J2000 coordinates listed in columns 4 and 5
  are of the BCG of the identified counterpart (Cen$=1$), or the centroid of the
  members (Cen$=2$). Columns 6, 7 and 8 show the cluster redshift, the
  spectroscopic redshift of the BCG (if available), and the number of galaxies
  with spectroscopic measurements. }
\label{tab:inpsz1}
\nointerlineskip
\vskip -3mm
\scriptsize
\setbox\tablebox=\vbox{
   \newdimen\digitwidth
   \setbox0=\hbox{\rm 0}
   \digitwidth=\wd0
   \catcode`*=\active
   \def*{\kern\digitwidth}
   \newdimen\signwidth
   \setbox0=\hbox{+}
   \signwidth=\wd0
   \catcode`!=\active
   \def!{\kern\signwidth}
\halign{\hfil#\hfil\tabskip=3em&
   \hbox to 1.5in{#\leaderfil}\tabskip=2em&
   \hfil#\hfil&
   \hfil#\hfil\tabskip=1em&
   \hfil#\hfil\tabskip=2em&
   \hfil#\hfil&
   \hfil#\hfil&
   \hfil#\hfil&
   #\hfil\tabskip=0pt\cr
\noalign{\doubleline}
\omit&\omit&\omit&\multispan2\hfil Position (J2000)\hfil\cr
\noalign{\vskip -3pt}
\omit&\omit&\omit&\multispan2\hrulefill\cr
ID&\omit\hfil \Planck\ Name\hfil&Cen.&R. A.&Decl.&$<z>$&$z_{\rm spec,BCG}$&$N_{\rm spec}$&\omit\hfil Notes\hfil\cr
\noalign{\vskip 3pt\hrule\vskip 5pt}
*42\rlap{$^{\rm b}$}     & PSZ1 G018.09$+$26.07 & 1 &  16 52 12.3 &  $-$00 20 15.7 &  0.090* & 0.0893 &  17 &  \cr
*85\rlap{$^{\rm b,d}$}   & PSZ1 G031.91$+$67.94 & 2 &  14 30 23.3 &  $+$24 39 06.2 &  0.134* &  $-$   &  54 &  \cr 
116\rlap{$^{\rm b,e,i}$} & PSZ1 G042.33$+$17.46 & 1 &  18 04 16.1 &  $+$16 02 14.9 &  0.458* & 0.4574 &  *3 &  \cr
137\rlap{$^{\rm e}$}     & PSZ1 G046.13$+$30.75 & 1 &  17 17 05.6 &  $+$24 04 17.1 &  0.569* & 0.5673 &  13 &  \cr
141\rlap{$^{\rm d}$}     & PSZ1 G046.98$+$66.62 & 1 &  14 37 40.3 &  $+$30 12 00.3 &  0.340* & 0.3398 &  *3 &  WHL J143740.3$+$301200 \cr
174\rlap{$^{\rm e,h}$}   & PSZ1 G055.72$+$17.58 & 1 &  18 25 37.1 &  $+$27 42 55.4 &  0.195* & 0.1944 &  24 &  \cr 
193\rlap{$^{\rm h}$}     & PSZ1 G058.82$-$49.66 & 1 &  22 28 19.1 &  $-$05 34 41.7 &  0.595* & 0.5954 &  *5 &  \cr
200\rlap{$^{\rm b,e}$}   & PSZ1 G060.12$+$11.42 & 1 &  18 58 46.0 &  $+$29 15 33.6 &  0.223* & 0.224* &  13 &  \cr
207\rlap{$^{\rm b}$}     & PSZ1 G063.80$+$11.42 & 1 &  19 05 57.4 &  $+$32 32 53.5 &  0.428* & 0.426* &  12 &  \cr
217\rlap{$^{\rm b,e}$}   & PSZ1 G066.01$-$23.30 & 1 &  21 19 26.2 &  $+$15 21 06.4 &  0.249* & 0.2487 &  *6 &  \cr  
218                      & PSZ1 G066.20$+$12.87 & 1 &  19 04 15.4 &  $+$35 16 00.1 &  0.245* & 0.2462 &  10 &  \cr
220\rlap{$^{\rm e}$}     & PSZ1 G066.41$+$27.03 & 1 &  17 56 52.7 &  $+$40 08 06.1 &  0.574* & 0.5748 &  *2 &  WHL J269.219$+$40.13 \cr
236\rlap{$^{\rm b,d,e}$} & PSZ1 G071.57$-$37.96 & 1 &  22 17 15.8 &  $+$09 03 10.1 &  0.288* & 0.2882 &  *1 &  ACO 2429 \cr   
245                      & PSZ1 G073.64$+$36.49 & 1 &  17 09 35.6 &  $+$47 31 53.9 &  0.549* & 0.5496 &  *6 &  \cr
249\rlap{$^{\rm a,f}$}   & PSZ1 G074.75$-$24.59 & 1 &  21 45 55.1 &  $+$20 28 18.8 &  0.33** &  $-$   &  *0 &  ZwCl 2143.5+2014 \cr
252\rlap{$^{\rm e,f}$}   & PSZ1 G076.44$+$23.53 & 1 &  18 28 21.4 &  $+$48 04 28.4 &  0.168* & 0.168* &  21 &  \cr
257\rlap{$^{\rm a,f}$}   & PSZ1 G078.67$+$20.06 & 1 &  18 52 02.7 &  $+$49 01 17.8 &  0.49** &  $-$   &  *0 &  \cr
265\rlap{$^{\rm a,f}$}   & PSZ1 G080.62$-$46.81 & 1 &  23 00 59.6 &  $+$06 44 58.2 &  0.05** &  $-$   &  *0 &  RBS 1929 \cr
281\rlap{$^{\rm b,e}$}   & PSZ1 G084.41$-$12.43 & 1 &  21 37 46.6 &  $+$35 35 51.0 &  0.274* & 0.2741 &  24 &  \cr
283\rlap{$^{\rm b}$}     & PSZ1 G084.62$-$15.86 & 1 &  21 49 40.6 &  $+$33 10 34.6 &  0.364* & 0.3673 &  *2 &  \cr
287                      & PSZ1 G084.85$+$20.63 & 1 &  19 00 11.6 &  $+$54 42 14.5 &  0.367* & 0.367* &  *8 &  \cr
297\rlap{$^{\rm b,d}$}   & PSZ1 G087.25$-$41.86 & 1 &  23 05 18.3 &  $+$13 34 36.1 &  0.0481 & 0.0481 &  *3 &  \cr  
312\rlap{$^{\rm c}$}     & PSZ1 G091.82$+$26.11 & 2 &  18 31 12.0 &  $+$62 14 50.1 &  0.822* &  $-$   &  11 &  \cr
316\rlap{$^{\rm b,d,i}$} & PSZ1 G092.41$-$37.39 & 2 &  23 10 14.7 &  $+$19 21 41.0 &  0.114* &  $-$   &  *3 &  \cr     
334\rlap{$^{\rm a,f}$}   & PSZ1 G095.49$+$16.41 & 1 &  20 00 06.7 &  $+$62 26 39.9 &  0.37** &  $-$   &  *0 &  \cr
337\rlap{$^{\rm a,f}$}   & PSZ1 G096.89$+$24.17 & 1 &  18 56 41.3 &  $+$66 21 56.0 &  0.37** &  $-$   &  *0 &  ZwCl 1856.8+6616 \cr
342\rlap{$^{\rm c}$}     & PSZ1 G097.93$+$19.46 & 1 &  19 44 24.0 &  $+$65 52 23.9 &  0.248* & 0.2454 &  *4 &  4C 65.28 \cr
359\rlap{$^{\rm e}$}     & PSZ1 G100.18$-$29.68 & 1 &  23 21 00.8 &  $+$29 12 09.7 &  0.485* &  $-$   &  *4 &  WHL J232104.1$+$291134 \cr
367\rlap{$^{\rm b}$}     & PSZ1 G102.86$-$31.07 & 1 &  23 33 24.7 &  $+$28 43 19.9 &  0.5915 & 0.5915 &  *1 &  \cr
373                      & PSZ1 G103.58$+$24.78 & 1 &  19 04 54.2 &  $+$72 28 33.1 &  0.334* &  $-$   &  *3 &  \cr
380\rlap{$^{\rm g}$}     & PSZ1 G105.91$-$38.39 & 0 &	  $-$	  &	 $-$	   &   $-$   &  $-$   &  *0 &  \cr
393\rlap{$^{\rm b}$}     & PSZ1 G108.18$-$11.53 & 2 &  23 22 27.2 &  $+$48 44 19.3 &  0.335* &  $-$   &  *6 &  \cr
395\rlap{$^{\rm a,f}$}   & PSZ1 G108.52$+$32.30 & 2 &  17 08 21.6 &  $+$76 30 06.3 &  0.26** &  $-$   &  *0 &  \cr
423\rlap{$^{\rm g}$}     & PSZ1 G115.34$-$54.89 & 0 &	  $-$	  &	 $-$	   &   $-$   &  $-$   &  *0 &  \cr
424\rlap{$^{\rm b,d}$}   & PSZ1 G115.59$-$44.47 & 1 &  00 29 06.3 &  $+$18 08 56.8 &  0.167* & 0.167  &  *1 &  \cr
425\rlap{$^{\rm b,e}$}   & PSZ1 G115.70$+$17.51 & 1 &  22 27 06.8 &  $+$78 20 08.1 &  0.111* & 0.111  &  *5 &  \cr
431\rlap{$^{\rm b}$}     & PSZ1 G118.06$+$31.10 & 1 &  15 54 38.7 &  $+$84 10 28.4 &  0.195* & 0.194  &  16 &  \cr
450\rlap{$^{\rm e,f}$}   & PSZ1 G123.55$-$10.34 & 1 &  00 55 24.5 &  $+$52 29 20.2 &  0.106* & 0.1064 &  30 &  \cr
465\rlap{$^{\rm g}$}     & PSZ1 G127.55$+$20.84 & 0 &	  $-$	  &	 $-$	   &   $-$   &  $-$   &  *0 &  \cr
470                      & PSZ1 G130.15$-$17.01 & 1 &  01 30 52.8 &  $+$45 17 46.7 &  0.211* & 0.2097 &  20 &  \cr
500\rlap{$^{\rm b}$}     & PSZ1 G138.60$-$10.85 & 1 &  02 27 06.7 &  $+$49 00 29.3 &  0.700* & 0.7023 &  *4 &  \cr
502\rlap{$^{\rm e}$}     & PSZ1 G139.61$+$24.20 & 1 &  06 21 49.3 &  $+$74 42 05.4 &  0.266* & 0.265  &  22 &  \cr
521\rlap{$^{\rm b,d}$}   & PSZ1 G146.00$-$49.42 & 2 &  01 51 19.8 &  $+$10 47 30.0 &  0.097* &  $-$   &  *4 &  WHL J015128.7$+$104912\cr 
541\rlap{$^{\rm b,h}$}   & PSZ1 G153.41$+$36.58 & 2 &  08 42 42.9 &  $+$62 30 21.6 &  0.650* &  $-$   &  *3 &  \cr
542\rlap{$^{\rm b,d}$}   & PSZ1 G153.56$+$36.23 & 2 &  08 39 33.4 &  $+$62 26 12.1 &  0.132* &  $-$   &  *6 &  MaxBCG J129.79086$+$62.44628 \cr
543\rlap{$^{\rm b,d}$}   & PSZ1 G153.87$+$41.05 & 1 &  09 18 21.4 &  $+$61 10 29.3 &  0.279* & 0.2794 &  *4 &  WHL J091821.4+611029 \cr
545\rlap{$^{\rm c}$}     & PSZ1 G155.25$-$68.42 & 1 &  01 37 24.9 &  $-$08 27 22.7 &  0.566* & 0.5661 &  *5 &  WHL J24.3324$-$8.477 \cr
574\rlap{$^{\rm b,d,h}$} & PSZ1 G165.76$+$31.15 & 1 &  07 57 58.7 &  $+$52 38 16.7 &  0.259* & 0.2593 &  *6 &  MaxBCG J119.49440$+$52.63797 \cr  
575\rlap{$^{\rm d}$}     & PSZ1 G165.94$+$50.48 & 1 &  10 01 38.8 &  $+$50 00 54.0 &  0.172* & 0.1709 &  *2 &  GMBCG J150.41158$+$50.01489 \cr
579\rlap{$^{\rm g}$}     & PSZ1 G167.43$-$38.04 & 0 &	  $-$	  &	 $-$	   &   $-$   &  $-$   &  *0 &  \cr
590\rlap{$^{\rm f}$}     & PSZ1 G171.96$-$40.64 & 1 &  03 12 56.9 &  $+$08 22 18.0 &  0.272* & 0.2716 &  16 &  \cr
594\rlap{$^{\rm e}$}     & PSZ1 G172.93$+$21.31 & 1 &  07 07 38.1 &  $+$44 19 54.3 &  0.336* & 0.339  &  17 &  \cr
615\rlap{$^{\rm b,d}$}   & PSZ1 G182.49$-$57.09 & 1 &  02 43 13.9 &  $-$08 07 30.3 &  0.031* & 0.031  &  37 &  \cr
629\rlap{$^{\rm c}$}     & PSZ1 G186.81$+$07.31 & 1 &  06 29 21.8 &  $+$26 30 19.4 &  0.221* & 0      &  20 &  WHL J97.3409$+$26.50 \cr
649\rlap{$^{\rm c}$}     & PSZ1 G194.24$+$75.22 & 1 &  11 46 17.3 &  $+$30 56 18.5 &  0.388* & 0.3927 &  *2 &  \cr
650\rlap{$^{\rm g}$}     & PSZ1 G194.68$-$49.73 & 0 &	  $-$	  &	 $-$	   &   $-$   &  $-$   &  *0 &  \cr
665\rlap{$^{\rm a,f}$}   & PSZ1 G200.95$-$28.16 & 1 &  04 50 24.3 &  $-$02 58 36.7 &  0.27** &  $-$   &  *0 &  \cr
672\rlap{$^{\rm f}$}     & PSZ1 G204.07$+$16.51 & 1 &  07 35 47.4 &  $+$15 06 49.2 &  0.121* & 0.1220 &  22 &  ZwCl 0733.1$+$1514 \cr
673\rlap{$^{\rm c}$}     & PSZ1 G204.24$+$14.51 & 1 &  07 28 35.4 &  $+$14 07 42.1 &  0.349* & 0.3484 &  *5 &  WHL J112.147$+$14.12 \cr
674                      & PSZ1 G204.73$+$15.85 & 1 &  07 34 27.8 &  $+$14 16 39.3 &  0.347* & 0.3471 &  16 &  \cr
682\rlap{$^{\rm b}$}     & PSZ1 G206.52$-$26.37 & 1 &  05 06 17.2 &  $-$06 34 52.5 &  0.391* & 0.391* &  *1 &  \cr
689\rlap{$^{\rm e,f}$}   & PSZ1 G209.80$+$10.23 & 1 &  07 22 23.7 &  $+$07 24 30.3 &  0.678* & 0.6780 &  *5 &  \cr
700\rlap{$^{\rm c,d,h}$} & PSZ1 G213.27$+$78.35 & 1 &  11 59 22.0 &  $+$26 27 04.0 &  0.267* & 0.2667 &  *1 &  GMBCG J179.84162$+$26.45111\cr
791\rlap{$^{\rm b,d}$}   & PSZ1 G236.51$+$48.39 & 2 &  10 20 47.2 &  $+$06 04 50.4 &  0.112* &  $-$   &  *7 &  \cr
807                      & PSZ1 G240.39$+$07.91 & 1 &  08 13 55.5 &  $-$20 06 56.0 &  0.332* & 0.3315 &  *3 &  \cr
844\rlap{$^{\rm b,d}$}   & PSZ1 G249.01$+$73.75 & 2 &  11 57 05.6 &  $+$16 57 35.4 &  0.156* &  $-$   &  *2 &  WHL J115659.6$+$165833 \cr
908\rlap{$^{\rm b,d}$}   & PSZ1 G264.01$+$59.77 & 1 &  11 37 53.5 &  $+$02 38 55.5 &  0.141* & 0.1409 &  *2 &  \cr
\noalign{\vskip 3pt\hrule\vskip 3pt}}}
\endPlancktablewide
\tablenote {{\rm a}} Redshift is estimated from our photometric data.\par
\tablenote {{\rm b}} Redshift value not published in the PSZ1 \citep{planck2013-p05a}\par
\tablenote {{\rm c}} Spectroscopic redshift. The redshift given in the PSZ1 catalogue \citep{planck2013-p05a} is photometric.\par
\tablenote {{\rm d}} Redshift estimated from SDSS data. \par
\tablenote {{\rm e}} Target also included in the RTT150 paper \citep{planck2014-XXVI}.\par
\tablenote {{\rm f}} PSZ1 containts a spectroscopic redshift obtained with other facility \citep{planck2013-p05a}.\par
\tablenote {{\rm g}} No optical counterpart identified.\par
\tablenote {{\rm h}} Several counterparts identified.\par
\tablenote {{\rm i}} Target also observed in the $r$ and $z$ bands with MegaCam at CFHT \citep{megacam}. \par
\endgroup
\end{table*}

\begin{table*}
\begingroup
\caption{Confirmed clusters not in the PSZ1 catalogue observed for this paper.} 
\label{tab:nopsz1}
\nointerlineskip
\vskip -3mm
\footnotesize
\setbox\tablebox=\vbox{
   \newdimen\digitwidth
   \setbox0=\hbox{\rm 0}
   \digitwidth=\wd0
   \catcode`*=\active
   \def*{\kern\digitwidth}
   \newdimen\signwidth
   \setbox0=\hbox{+}
   \signwidth=\wd0
   \catcode`!=\active
   \def!{\kern\signwidth}
\halign{\hbox to 1.5in{#\leaderfil}\tabskip=2em&
   \hfil#\hfil&
   \hfil#\hfil\tabskip=1em&
   \hfil#\hfil\tabskip=2em&
   \hfil#\hfil&
   \hfil#\hfil&
   \hfil#\hfil&
   \hfil#\hfil\tabskip=0pt\cr
\noalign{\doubleline}
\omit&\omit&\multispan2\hfil Position (J2000)\hfil\cr
\noalign{\vskip -3pt}
\omit&\omit&\multispan2\hrulefill\cr
\omit\hfil Name\hfil&Cen.&R. A.&Decl.&$<z>$&$z_{\rm spec,BCG}$&N$_{\rm spec}$&Notes\cr
\noalign{\vskip 3pt\hrule\vskip 5pt}
PLCK G081.38$-$60.03& 1 & 23 34 03.6 & $-$03 35 55.2 & 0.260* & 0.2596 & 11 & \cr         
PLCK G087.67$+$22.99& 1 & 18 48 56.6 & $+$57 53 29.5 & 0.33** & $-$    & *0 & {\rm a}\cr  
PLCK G100.24$-$30.45& 1 & 23 22 14.7 & $+$28 31 15.0 & 0.394* & 0.404* & 12 & {\rm b}\cr  
PLCK G107.02$+$37.62& 1 & 16 02 24.6 & $+$72 57 29.7 & 0.33** & $-$    & *0 & {\rm a}\cr  
PLCK G115.12$+$28.56& 1 & 17 57 36.1 & $+$82 56 33.9 & 0.169* & 0.1684 & 31 & \cr        
PLCK G128.38$+$71.18& 1 & 12 41 58.8 & $+$45 46 47.5 & 0.336* & 0.3367 & 12 & {\rm c}\cr 
PLCK G134.25$-$44.24& 1 & 01 25 47.9 & $+$17 59 59.4 & 0.232* & 0.2311 & *4 & {\rm c}\cr  
PLCK G164.50$-$24.92& 1 & 03 36 35.9 & $+$24 20 45.0 & 0.35** & $-$    & *0 & {\rm a}\cr  
PLCK G203.23$+$31.93& 1 & 08 35 27.4 & $+$21 36 32.2 & 0.1349 & 0.1349 & *1 & {\rm d}\cr  
PLCK G217.69$+$67.29& 1 & 11 13 26.3 & $+$23 11 04.6 & 0.2454 & 0.2454 & *1 & {\rm d}\cr  
PLCK G225.44$+$51.89& 2 & 10 17 52.0 & $+$13 56 04.0 & 0.297* & $-$    & *3 & {\rm c}\cr  
\noalign{\vskip 3pt\hrule\vskip 3pt}}}
\endPlancktablewide
\tablenote {{\rm a}} Redshift is estimated from our photometric data.\par
\tablenote {{\rm b}} Target included in XMM sample \citep{planck2011-5.1b}. \par
\tablenote {{\rm c}} Several possible counterparts identified. \par
\tablenote {{\rm d}} Spectroscopic redshift estimated from SDSS data. \par
\endgroup
\end{table*}

\begin{figure}[ht!]
\centering
\includegraphics[width=\columnwidth]{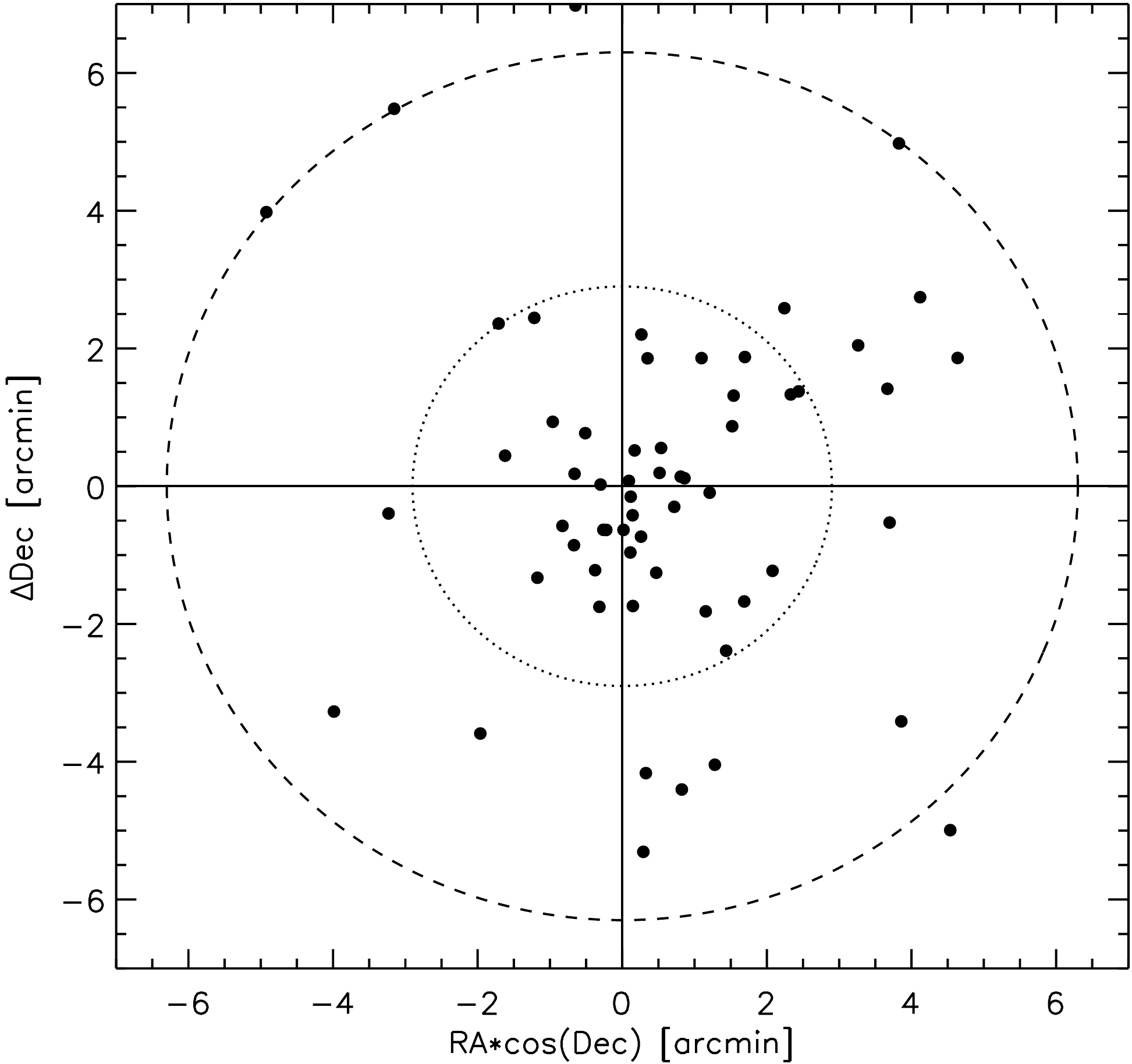}
\caption{Distribution of cluster optical centre offsets relative to their
  \Planck\ SZ position. The dotted (dashed) line corresponds to a radius of
  $2\parcm9$ ($6\parcm3$), which encloses 68\% (95\%) of the full PSZ1 sample
  presented in this paper.}
\label{fig:offsets}
\end{figure}

\begin{figure}[ht!]
\centering
\includegraphics[width=\columnwidth]{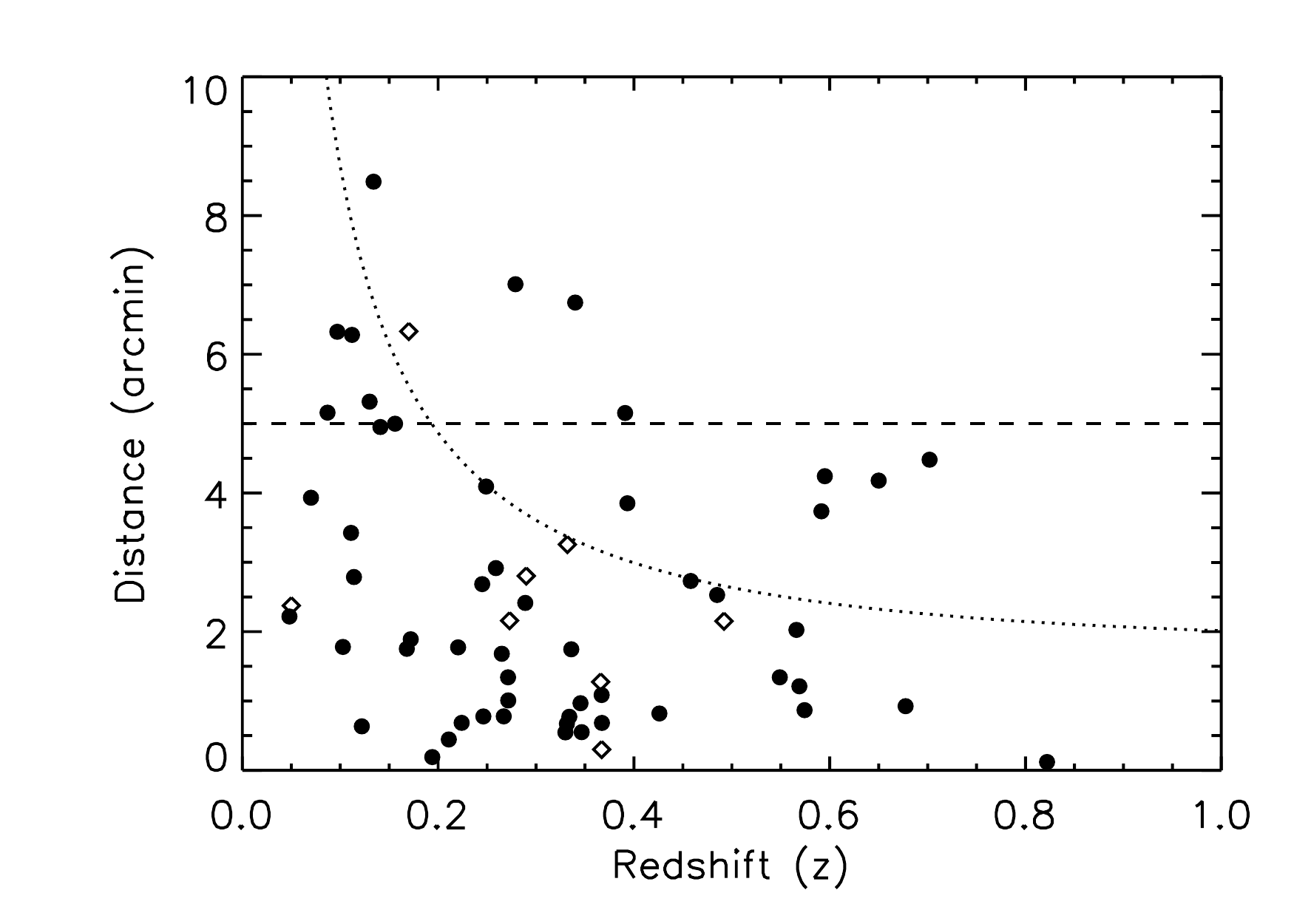}
\caption{ Cluster optical centre offsets relative to their \Planck\ SZ position
  as a function of cluster redshift. Open symbols correspond to photometric
  redshift determinations; filled dots are spectroscopic measurements. The
  dashed horizontal line at $5\arcm$ shows the maximum offset expected for a
  \Planck\ SZ detection (i.e. a \Planck\ beam). The dotted line corresponds to
  the angle subtended by 1\,Mpc in projection at the corresponding redshift. }
\label{fig:d-vs-z}
\end{figure}

Tables \ref{tab:inpsz1} and \ref{tab:nopsz1} provide the coordinates of the
cluster optical centres calculated from the position of the BCG. If a BCG is not
clearly identified or if two BCGs are found, the coordinates refer to the
(approximate) geometrical centre of the cluster members. The distribution of the
optical centre offsets relative to the SZ positions measured by \Planck\ is
shown in Fig.~\ref{fig:offsets}. Sixty-eight percent of the sample have a
positional error smaller than $2\parcm9$, consistent with (although slightly larger
than) the expectation for the whole PSZ1 of $2\arcm$ \citep{planck2013-p05a}.
Figure~\ref{fig:d-vs-z} shows the distance of the cluster optical centres from
the nominal \Planck\ position as a function of redshift. For low-redshift
systems ($0.15\lsim z \lsim 0.20$), the characteristic scale of the virial
radius of a cluster (here taken to be 1\,Mpc) extends beyond $5\arcm$. In
principle, associations between \Planck\ SZ candidates and low-redshift optical
counterparts with large apparent angular separations might therefore be
expected.  When we restrict the analysis in Fig.~\ref{fig:offsets} to the
sub-sample of clusters with redshifts above 0.2, the positional error is reduced
to $2\parcm5$.

\begin{figure}[ht!]
\centering
\includegraphics[width=\columnwidth]{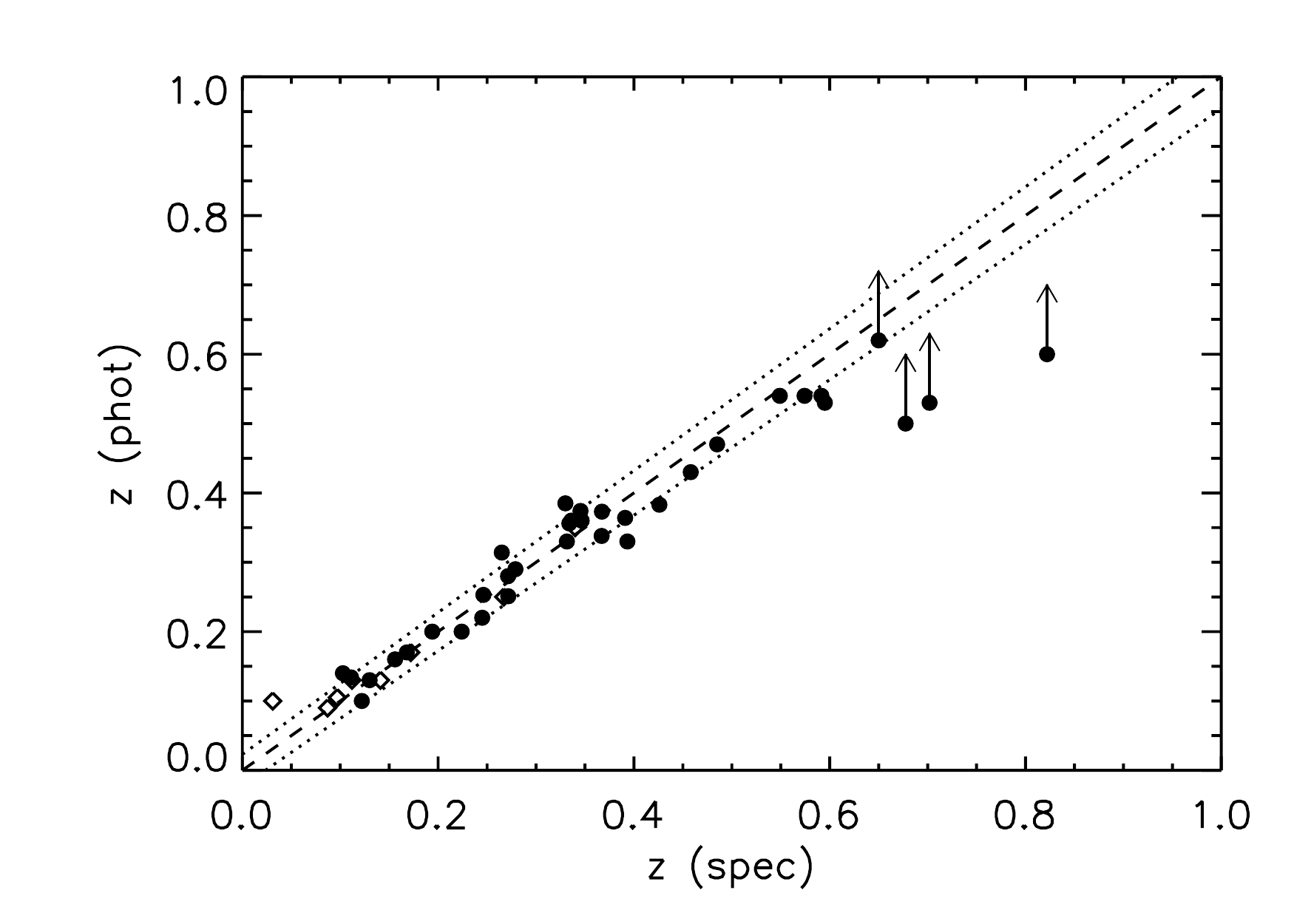}
\caption{ Comparison of our photometric redshift estimates based on red-sequence
  colours with spectroscopic redshifts. The statistical error for the full
  sample is about 2.5\,\%. For completeness, we also include as open symbols
  the few cases in which the photometric redshifts are derived from the SDSS
  database. }
\label{fig:zphot}
\end{figure}

Figure~\ref{fig:zphot} shows the comparison of our photometric redshift
estimates (based on the method described in Sect.~\ref{sec:photoz}) with
the corresponding spectroscopic redshifts for all targets in the PSZ1 catalogue
described in Table~\ref{tab:inpsz1}. The scatter between the two estimates is
$\delta z/(1+z) = 0.025$.

We note that all sky images presented in this section are oriented with north up and east to the left.

\subsection{Notes on individual objects from the PSZ1 catalogue}
\label{sec:notes}

\paragraph{PSZ1 G031.91$+$67.94} There is no obvious identification
within a distance of $5\arcm$ from the \Planck\ SZ centre; however, the field is
very rich in the surroundings. The \citet{Tempel2012} catalogue cited a few poor
galaxy groups (29401, 29403, 29404, 28941, and 29795) and two large galaxy
clusters (29405 at $z=0.134$; 28941 at $z=0.039$) within a distance of
approximately $10\arcm$ from the \Planck\ centre. Inspection of the Compton
$y$-map in this area shows that none of the BCGs of these clusters corresponds
to the main peak of the SZ emission. By inspecting a large area of about one
square degree in the optical catalogs, we identify a clear peak in the galaxy
density at $18\arcm$ from the \Planck\ centre, which we list as the optical
counterpart in Table~\ref{tab:inpsz1}. Using the current version of SDSS DR12,
we found 54 spectroscopic members for this cluster over a region larger than
$30\arcm$ at redshift $z = 0.134$ with $\sigma_{\rm v} \approx
650$\,km\,s$^{-1}$. There is no obvious BCG for this system, and it shows
several galaxy clumps (possibly substructures) in the 2D galaxy distribution,
each clump showing bright galaxies. A detailed dynamical study of this complex
cluster is needed, and a more detailed study of the SZ emission, in combination
with X-ray images, is required to confirm the proposed association. We note that
one of the clumps of this cluster seems to be associated with GMBCG
J217.49013$+$24.69973 \citep{hao2010}, with $\zphot = 0.135$.

\paragraph{PSZ1 G046.98$+$66.62} The proposed counterpart corresponds to
the cluster WHL J143740.3$+$301200 \citep{WHL2012}, with $\zphot=0.33$. Our
spectroscopic redshift is determined from SDSS data. Although the cluster is
located $6\parcm7$ from the \Planck\ centre, the Compton $y$-map
\citep{planck2013-p05b} supports this association, as did \citet{Liu2014}.

\begin{figure}[ht!]
\centering \includegraphics[width=\columnwidth]{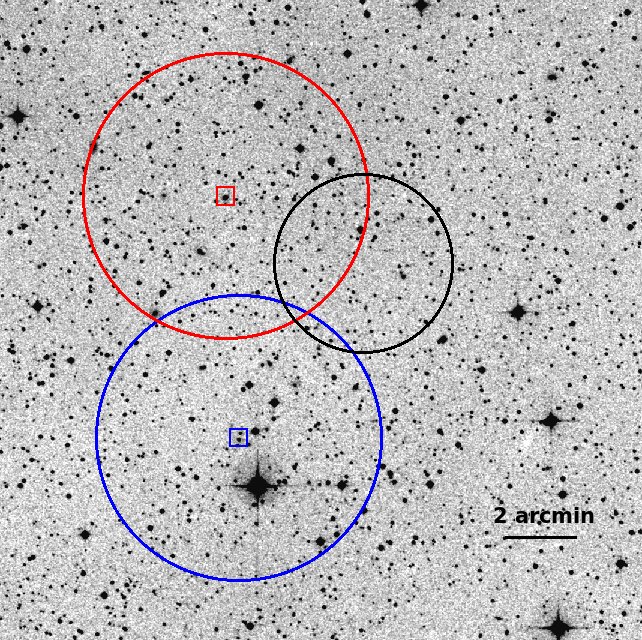}
\caption{ DSS infrared image for the PSZ1 G055.72$+$17.58 source. The black
  circle shows the location of the \Planck\ SZ centre with a $5\arcm$
  diameter. Red and blue circles correspond to our confirmed clusters at
  $z=0.070$ and $z=0.193$, with 17 and 24 galaxy members, respectively. Small
  squares mark the BCG for each system. The SZ signal seems to be mainly
  associated with the southern cluster (inside the blue circle). }
\label{fig:psz1-174} 
\end{figure}

\paragraph{PSZ1 G055.72$+$17.58} There are two possible identifications for this 
SZ source, both confirmed spectroscopically with the MOS at DOLORES/TNG (see
Fig.~\ref{fig:psz1-174} for details). The identification reported in
Table~\ref{tab:inpsz1} corresponds to a cluster with 24 spectroscopically
confirmed members at $z=0.193$. It is difficult to identify the BGC of this
cluster because there are only a few galaxies with comparable magnitudes, given our error
bars, therefore we provide the coordinates of the average position of all the
spectroscopically confirmed members. This target was also identified in
\citet{planck2014-XXVI} as the optical counterpart of the SZ source.  The
\Planck\ Compton $y$-map in this area \citep{planck2013-p05b} supports this
association.
We found a second cluster in the region, less massive than the previous cluster,
with 17 spectroscopically confirmed members at $z=0.070$.  Its BCG is at
($18^{\rm h}\ 25\arcm\ 38\parcs8$, $+27\deg\ 49\arcm\ 41\parcs6$), $3\parcm 9$
from the \Planck\ centre.

\paragraph{PSZ1 G058.82$-$49.66} The proposed counterpart is a rich cluster 
located approximately $4\arcm$ from the \Planck\ centre, with five spectroscopic
members obtained with OSIRIS/GTC. The spectroscopic redshift of the BCG is
$z=0.595$. This counterpart is well-aligned with the signal found in the Compton
$y$-map in this area \citep{planck2013-p05b}.
In addition, our RGB image obtained with the WFC/INT data shows another
overdensity of red galaxies, with photometric redshift $\zphot = 0.68$. However,
our OSIRIS/GTC observations in this second region are not conclusive. A
long-slit observation including the apparent BCG ($22^{\rm
  h}\ 27\arcm\ 59\parcs1$, $-5\deg\ 28\arcm\ 51\parcs6$) and two additional
(photometrically selected) galaxies of this second cluster give a redshift of
the BCG of $z = 0.7069\pm 0.0003$. Additional spectroscopic members need to be
confirmed before this can be firmly established as a galaxy cluster.

\begin{figure}[ht!]
\centering 
\includegraphics[width=\columnwidth]{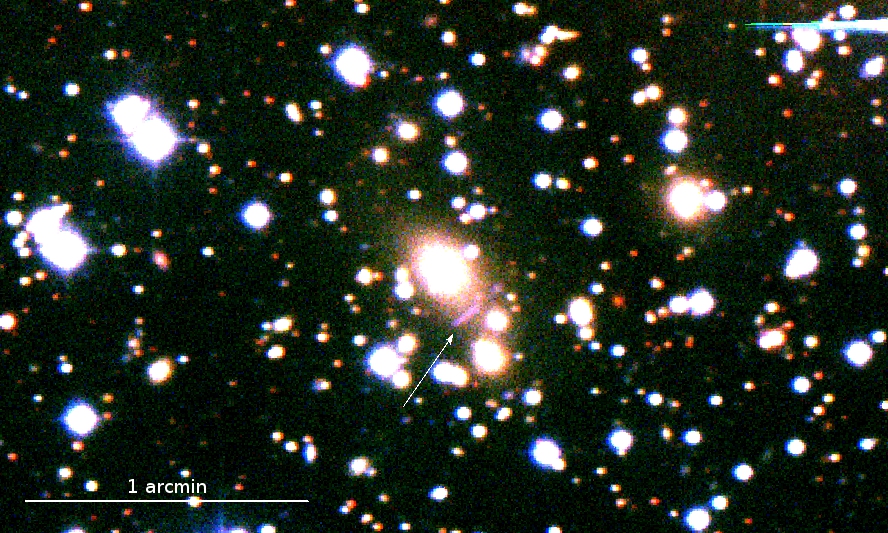}
\caption{ RGB image of the cluster PSZ1 G060.12$+$11.42, obtained with the
  WFC/INT data. The white arrow shows the location of a gravitational arc. }
\label{fig:psz1-200} 
\end{figure}

\paragraph{PSZ1 G060.12$+$11.42} There is a clear galaxy cluster 
at $\zphot = 0.20$. With DOLORES/TNG we confirm 18 spectroscopic members, and
the redshift of the BCG is $z=0.224$. There is also a clear arc seen in the RGB
image (see Fig.~\ref{fig:psz1-200}).

\paragraph{PSZ1 G076.44$+$23.53}  This is a clear fossil group
\citep[e.g.][]{jones2003,voevodkin2010}. It is difficult to estimate the
magnitude of the BGC because of the angular extension of the galaxy. Our
photometric redshift estimate, based on colour selection of cluster members, is
$z=0.17$. With DOLORES/TNG, we confirm 21 spectroscopic members, and the
redshift of the BCG is $z=0.168$. This source was also included in the RTT150
sample \citep{planck2014-XXVI}, and the spectroscopic redshift obtained there
for the BGC is fully consistent with our value.

\begin{figure}[ht!]
\centering
\includegraphics[width=\columnwidth]{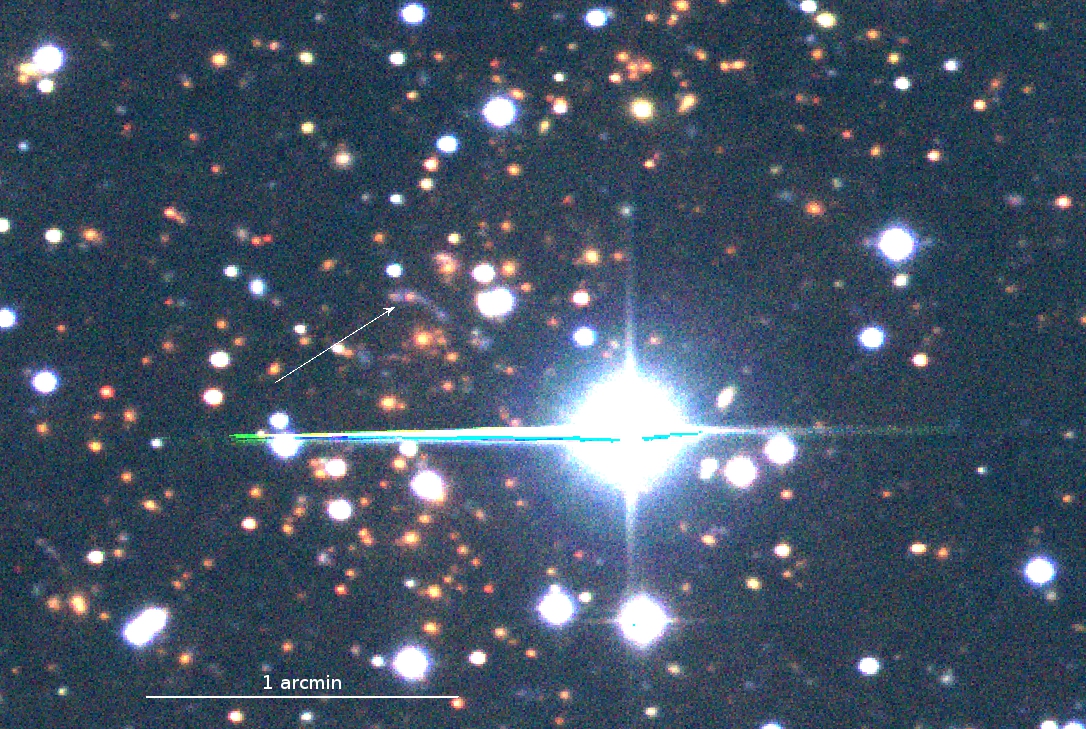}
\caption{ RGB image of the cluster PSZ1 G078.67$+$20.06 obtained with our
  WFC/INT data. A gravitational arc (indicated by the arrow) is clearly visible
  around the cluster centre. }
\label{fig:psz1-257}
\end{figure}

\paragraph{PSZ1 G078.67$+$20.06} Using WFC/INT, we identify a cluster with
  $\zphot=0.49$. The spectroscopic redshift for this cluster is $z=0.450$, as
obtained by the NOT telescope \citep{planck2013-p05a}. The optical image clearly
shows a gravitational arc around the BCG (see Fig.~\ref{fig:psz1-257}).

\paragraph{PSZ1 G080.62$-$46.81} This is identified in NED/SIMBAD as 
\object{RBS\,1929} (RX J2301.1$+$0646), with a spectroscopic redshift of
$z=0.042$. Our photometric analysis, combined with SDSS photometry, gives a
photometric redshift of $z=0.05$.
 
\paragraph{PSZ1 G084.85$+$20.63} Using DOLORES/TNG, we confirm nine
spectroscopic members for this cluster. The spectroscopic redshift provided in
Table~\ref{tab:inpsz1} of $z=0.367$ corresponds to the BCG. However, there is a
second galaxy with almost the same brightness in $r^\prime$ that might be
considered as a second BCG. The coordinates of this second galaxy are ($19^{\rm
  h}\ 00\arcm\ 03\parcs9$, $54\deg\ 42\arcm\ 37\parcs06$), and its redshift is
$z=0.375$.

\paragraph{PSZ1 G087.25$-$41.86} Fossil group, with a gap between the
BGC and the second cluster member of more than 2\,mag in $r^\prime$. The
spectroscopic redshift quoted in Table~\ref{tab:inpsz1} corresponds to the BCG
and was obtained from the SDSS database.

\paragraph{PSZ1 G091.82$+$26.11} The photometric redshift for this source was
incorrectly estimated in the first version of the PSZ1 catalogue
\citep{planck2013-p05a} ($\zphot=0.24$). Re-evaluation of the photometric
redshift suggested that this probably is a $z>0.60$ candidate, and this was
later confirmed with spectroscopic observations using the MOS mode at
OSIRIS/GTC. We have 18~spectrocopically confirmed members at $z=0.822$. This
cluster is part of the cosmological sample used in \citet{planck2013-p15}.

\paragraph{PSZ1 G092.41$-$37.39} This system is classified as a fossil group
with a double-nucleus BCG. The magnitude gap between the sum of the two fluxes
of the central galaxies and the other members is larger than 2\,mag in
$r^\prime$. The spectroscopic redshift was derived with OSIRIS/GTC. The cluster
richness estimated from MegaCam imaging data suggests a low-mass system, as is
also implied by the lack of any significant X-ray emission in the RASS maps
\citep[see also][]{megacam}.

\paragraph{PSZ1 G108.52$+$32.30} The optical counterpart is a poor system. 
Our photometric redshift estimate is $\zphot=0.26 \pm 0.02$. A spectroscopic
redshift with the NOT telescope of $z=0.250$ was included in the PSZ1 catalogue
and is fully consistent with our value.

\paragraph{PSZ1 G115.59$-$44.47}  
This target was included in the RTT150 sample \citep{planck2014-XXVI}, but no
counterpart was identified, in part due to possible contamination by Galactic
cirrus.  Inspection of the Compton $y$-map in this area \citep{planck2013-p05b}
shows an elongated structure with two peaks, but the detection might be
contaminated by an unresolved source visible in the 217\,GHz \Planck\ map. One
of the peaks corresponds to a possible galaxy cluster $6\parcm3$ from the
\Planck\ centre with a photometric redshift of $\zphot = 0.17$.  Firm
confirmation of this optical cluster counterpart will require spectroscopic
measurements.
%

\paragraph{PSZ1 G115.70+17.51} Our photometric redshift estimate for this
cluster, which is quite extended and has much substructure, was
$\zphot=0.13$. We performed MOS observations with DOLORES/TNG and found six
spectroscopic members. The BCG is located at $z=0.111$.
This source was discussed in detail in \citet{planck2014-XXVI}, where a second
possible red sequence of galaxies was identified in the background corresponding
to a redshift of $\zphot=0.5$. Slits were placed at the location of four of
these red galaxies in our MOS observations, but the derived redshifts are
not coincident, and therefore we do not confirm the presence of that second
cluster along the line of sight.

\begin{figure}[ht!]
  \centering
  \includegraphics[width=\columnwidth]{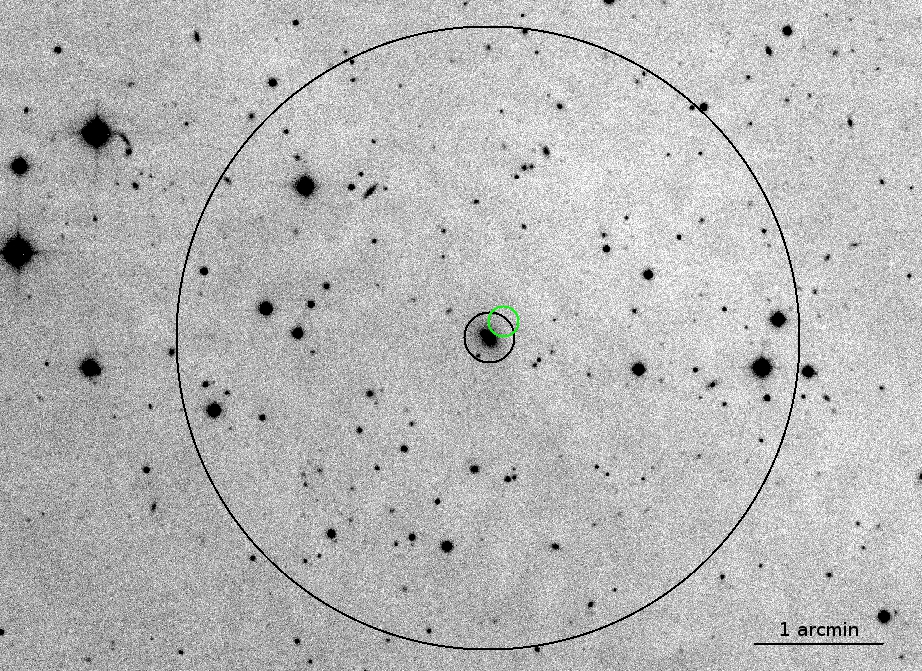}
  \includegraphics[width=\columnwidth]{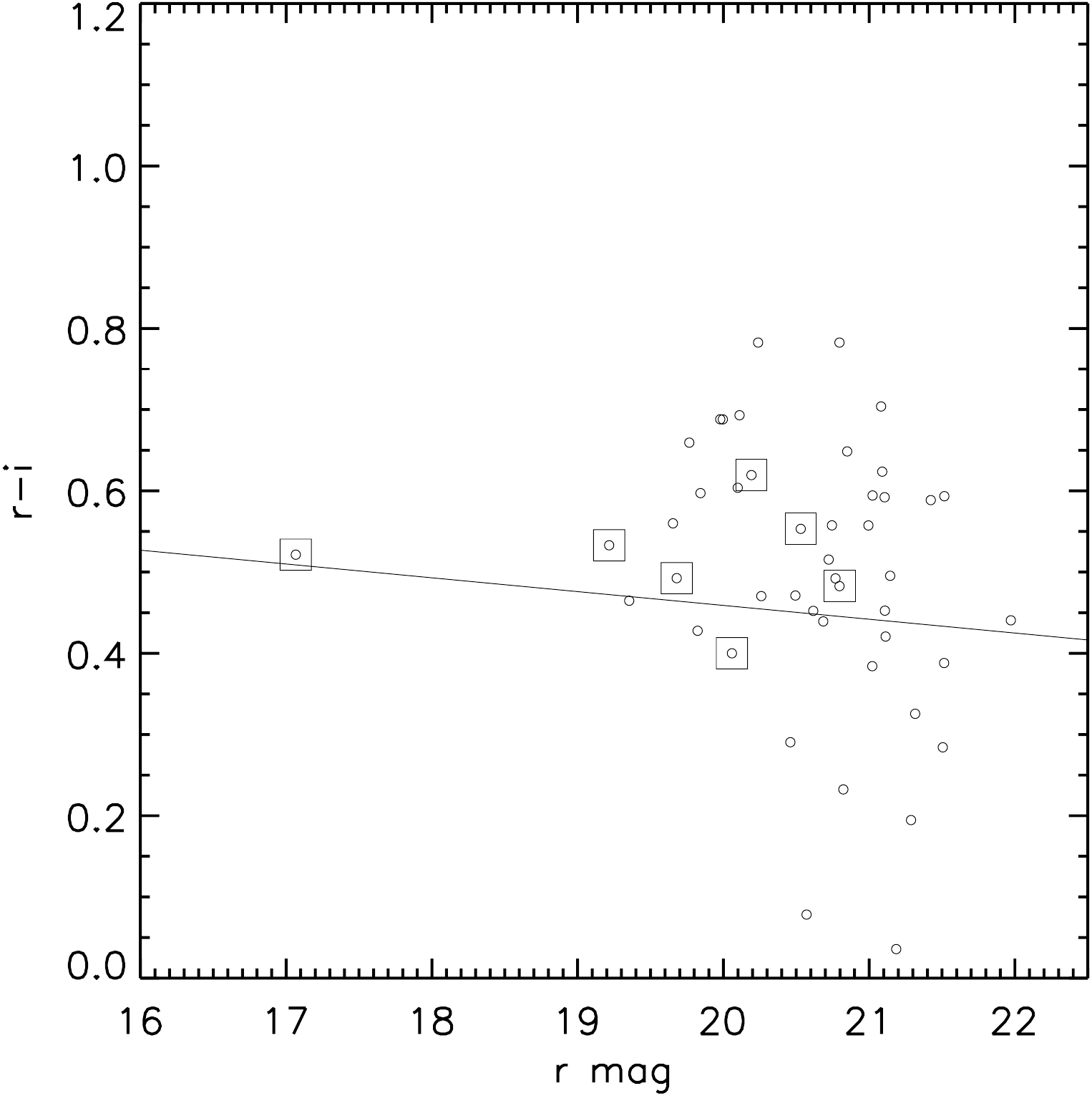}
  \caption{ PSZ1 G118.06$+$31.10, a galaxy cluster classified as fossil group
    following the definition of \citet{voevodkin2010}.\quad {\it Top\/}: WFC/INT
    $r^\prime$ band image of the cluster. The \Planck\ pointing is shown with a
    small green circle, and the large circle has $5\arcm$ diameter. The BCG is
    also highlighted.\quad {\it Bottom\/}: Colour-magnitude diagram of the
    galaxies near the cluster centre. The squares indicate the spectroscopically
    confirmed members within 0.5\,Mpc distance from the BCG.}
  \label{fig:psz1-431} 
\end{figure}

\paragraph{PSZ1 G118.06$+$31.10} Our photometric redshift estimate is
$\zphot=0.20$ (ACAM/WHT). The spectroscopic redshift of the BCG is $z=0.194$,
and we have 14 additional members spectroscopically confirmed with
DOLORES/TNG. This cluster was classified as a fossil group according to the
criteria of \cite{voevodkin2010} because the magnitude gap between the first and
second brightest galaxies is $\Delta m_{1,2} \approx 1.8$, within 0.5\,Mpc of
the BCG. Figure~\ref{fig:psz1-431} shows the colour-magnitude diagram for this
cluster. The red sequence in this case is given by $(r-i)=0.799-0.017r$
\citep{Barrena2012}.

\paragraph{PSZ1 G123.55$-$10.34} The optical counterpart is 
$1\parcm8$ from the \Planck\ centre. Our photometric redshift is $\zphot=0.14$
(WFC/INT) and the spectroscopic redshift of the BCG is $z=0.1027$. We have 27
spectroscopically confirmed members with DOLORES/TNG.  This is another fossil
group, as $\Delta m_{1,2} > 2$, although it is difficult to estimate the
magnitude of the BGC because of its angular extension. This target was also
included in the RTT150 paper \citep{planck2014-XXVI}, with a spectroscopic
redshift of $z=0.107$.

\begin{figure}[ht!]
\centering \includegraphics[width=\columnwidth]{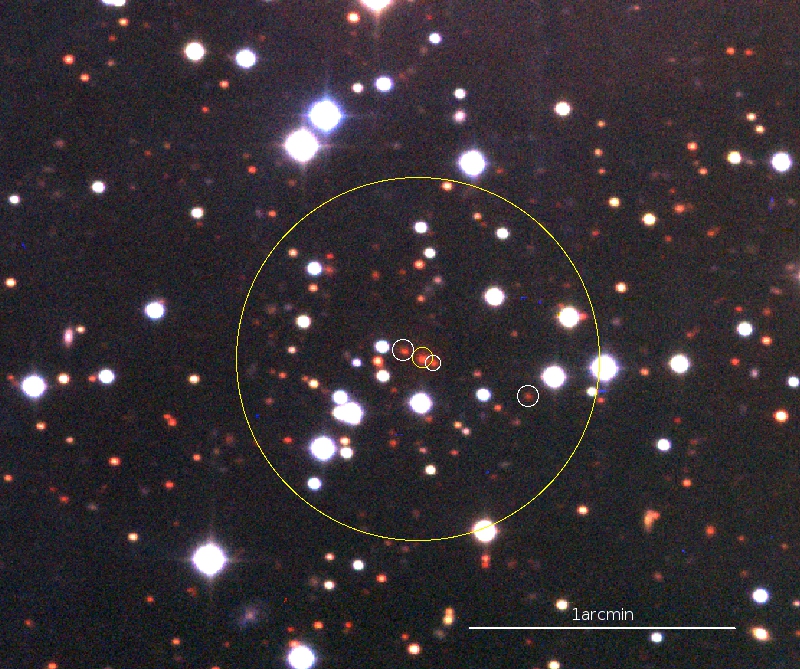}
\caption{ RGB image of PSZ1 G138.60$-$10.85, obtained with the WFC/INT. The
  small white circles show the location of our four spectroscopically confirmed
  galaxies at $z = 0.702$ with OSIRIS/GTC. The yellow circle indicates the
  location of the cluster, and its diameter corresponds to 0.5\,Mpc at that
  redshift. }
\label{fig:psz1-500}
\end{figure}

\paragraph{PSZ1 G138.60$-$10.85} Figure~\ref{fig:psz1-500} shows the RGB image
obtained with the WFC/INT. The optical counterpart for this SZ target is
$4\parcm5$ from the \Planck\ centre, and we obtained four spectroscopic members
with OSIRIS/GTC. The redshift of the BCG is $z=0.702$. As the galaxy members are
very red and we lacked a $Z$ -band image for this cluster, our original
photometric redshift determination was quite uncertain ($\zphot > 0.53$).

\paragraph{PSZ1 G146.00$-$49.42} Our WFC/INT image clearly shows Galactic cirrus
contamination in the region. The proposed optical counterpart for this cluster
is WHL J015128.7$+$104912, with $\zphot = 0.105$ \citep{WHL2012}. Here we
provide the spectroscopic confirmation, with four members at the same redshift
from the SDSS database. The quoted value corresponds to the spectroscopic
redshift of the BCG, $z=0.097$. However, we note that the proposed BCG is a disk
galaxy, showing clear signs of star formation (\ion{O}{II}, \ion{H}{$\alpha$},
and \ion{N}{II} emission lines). In addition, the magnitude gap among the three
brightest galaxies is quite small ($\Delta m_{1,2}=0.2$ and $\Delta
m_{2,3}=0.2$). We therefore provide for this case as the central coordinates of
the cluster the centroid of the positions of ten photometrically selected
members. This target was also observed with the RTT150 telescope
\citep{planck2014-XXVI}, but no counterpart was identified.

\begin{figure}[ht!]
\centering \includegraphics[width=\columnwidth]{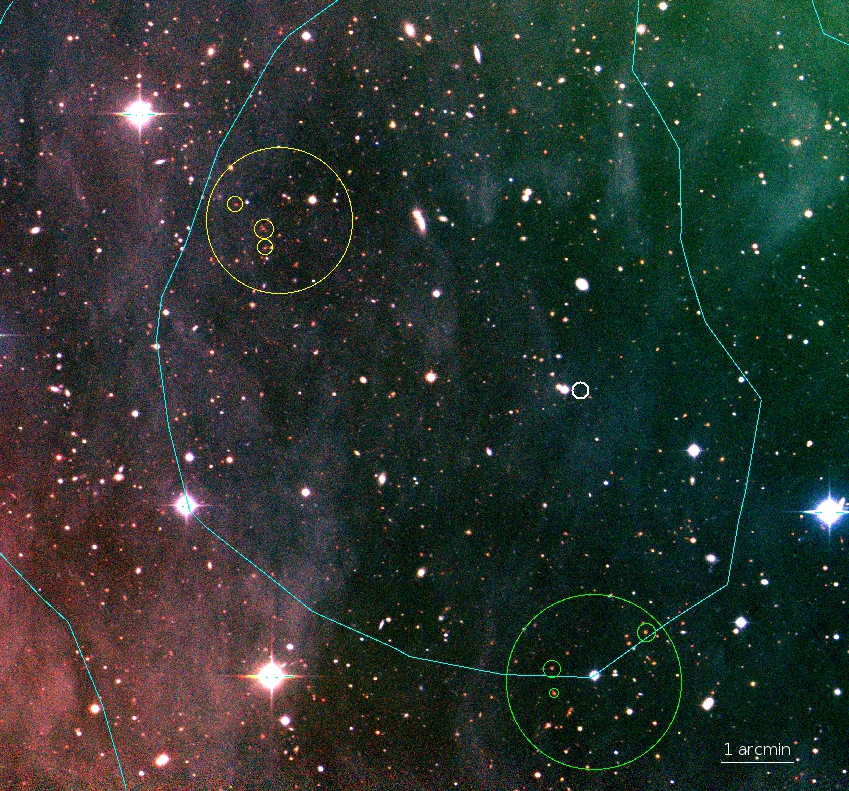}
\caption{RGB image of the cluster PSZ1 G153.41$+$36.58. The white circle shows
  the location of the \Planck\ SZ centre. Small circles correspond to the
  locations of the spectroscopically confirmed clusters at $z=0.65$ (green) and
  $z=0.825$ (yellow). The light blue contours correspond to the $2.5\times
  10^{-6}$ and $4\times 10^{-6}$ levels of the Compton $y$-map in this area
  \citep{planck2013-p05b}.  }
\label{fig:psz1-541}
\end{figure}

\paragraph{PSZ1 G153.41$+$36.58} There are two possible counterparts
associated with this \Planck\ SZ source (see Fig.~\ref{fig:psz1-541}). The first
is a cluster spectroscopically confirmed at $z=0.650$ (three spectroscopic
members) with OSIRIS/GTC, at a distance of about $4\arcm$ from the
\Planck\ position. However, there is a second cluster with three galaxies
spectroscopically confirmed at $z=0.825$, but $4.9\arcm$ from the
\Planck\ position. In Table~\ref{tab:inpsz1} we report as the optical
counterpart the cluster that is closer to the \Planck\ position; however,
inspection of the \Planck\ Compton $y$-map in this area \citep{planck2013-p05b}
shows an elongated structure that encompasses both clusters.

\paragraph{PSZ1 G153.56$+$36.23} Using the SDSS database, we
identified a low-redshift counterpart for this cluster, $5\parcm3$ from the
nominal \Planck\ position, with six spectroscopic members, all of them around
$z=0.130$. There is no clear BCG for this cluster. The image shows three large
galaxies with similar $r$ magnitudes ($16.9$--$17.1$). The coordinates in
Table~\ref{tab:inpsz1} correspond to the geometrical centroid of all
photometrically selected members. A search in NED around this position yields
two galaxy clusters: MaxBCG J129.79086$+$62.44628 with photometric redshift
$\zphot=0.146$ \citep{koester2007b}, and GMBCG J129.95605+62.41001 with
$\zphot=0.168$ \citep{hao2010}. However, the proposed BCGs for these two
photometrically selected clusters are indeed at the same spectroscopic redshift
of $z=0.132$, and therefore we conclude that they are all part of the same
structure that we identify here.
This target is only $22\arcm$ away from PSZ1 G153.41$+$36.58.

\paragraph{PSZ1 G153.87$+$41.05} Using the SDSS database, we identify a 
counterpart for this cluster $7\arcm$ from the \Planck\ position, with a
spectroscopic redshift of $z=0.279$. Although this distance is unusually large,
the Compton $y$-map in this area \citep{planck2013-p05b} supports this
association.

\paragraph{PSZ1 G165.76$+$31.15} All our redshift estimates for this candidate 
are based on SDSS data (both photometry and spectroscopy). In this region, we
found two clusters overlapped in projection along the line of sight. The cluster
reported in Table~\ref{tab:inpsz1} corresponds to MaxBCG J119.49440$+$52.63797,
a photometrically selected cluster with $\zphot = 0.30$ \citep{koester2007b},
$2\parcm9$ from the \Planck\ position. Here we confirm this cluster
spectroscopically, and we identified six spectroscopic members using SDSS
data. The spectroscopic redshift of the central galaxy is $z=0.259$.
Nearby, we found a fossil group ($\Delta m_{1,2} \approx 2.5$) at $z=0.041$, with
13 spectroscopically confirmed members extracted from SDSS data. The BCG for
this second cluster is at $07^{\rm h}\ 57\arcm\ 31\parcs37$,
$+52\deg\ 40\arcm\ 58\parcs3$), but the galaxy members are located
throughout the
region. This structure is classified as a group in \citet{Tempel2014}, with
reference GroupID$=1889$.

\paragraph{PSZ1 G182.49$-$57.09} The proposed counterpart is a low-redshift
cluster, with the BCG at $z=0.031$, and with 37 spectroscopic members selected
from SDSS. The cluster is spread over an area of approximately 1\,deg$^2$, which
corresponds to a physical size of 2\,Mpc at that redshift.  The proposed BCG is
$20\arcm$ from the \Planck\ pointing.

\paragraph{PSZ1 G186.81$+$07.31} The SZ source was identified as
WHL J97.3409$+$26.50 \citep{WHL2012}, with photometric redshift
$\zphot=0.258$. Spectroscopic measurements of 18~members with the MOS at
DOLORES/TNG give a redshift of $z=0.2204$.

\begin{figure}[ht!]
\centering
\includegraphics[width=\columnwidth]{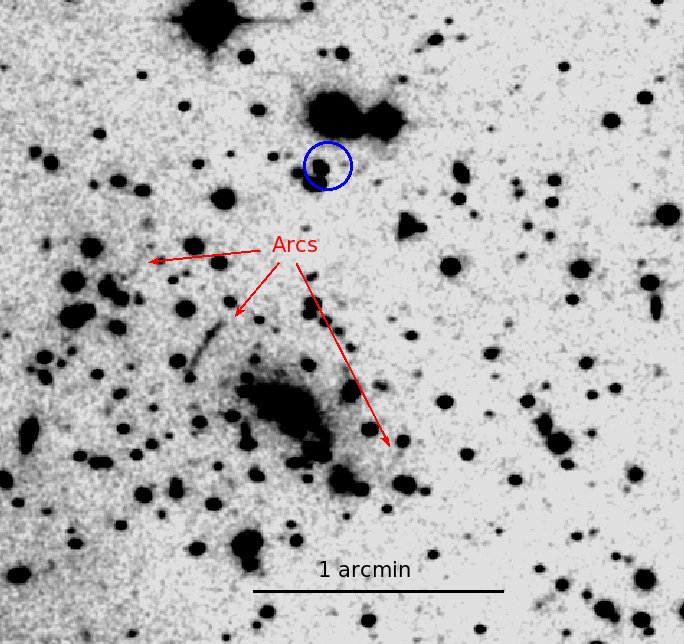}
\caption{WFC/INT $i^\prime$ band image of the cluster PSZ1 G204.24$+$14.51. The
  blue circle shows the \Planck\ PSZ1 position. Gravitational arcs (indicated by
  the red arrows) are clearly visible around the cluster centre.}
\label{fig:psz1-673}
\end{figure}

\paragraph{PSZ1 G204.24$+$14.51} This is a clear optical detection, identified
as WHL J112.147$+$14.12 with a published photometric redshift of $\zphot =
0.355$ \citep{WHL2012}. Here, we provide the spectroscopic confirmation of this
cluster, observing two confirmed cluster members with the NOT telescope, and
three additional members with the GTC. Our spectroscopic redshift for the BCG is
$z=0.3454$. Our optical image clearly shows several gravitational arcs around
the BCG (see Fig.~\ref{fig:psz1-673}).

\begin{figure}[ht!]
\centering \includegraphics[width=\columnwidth]{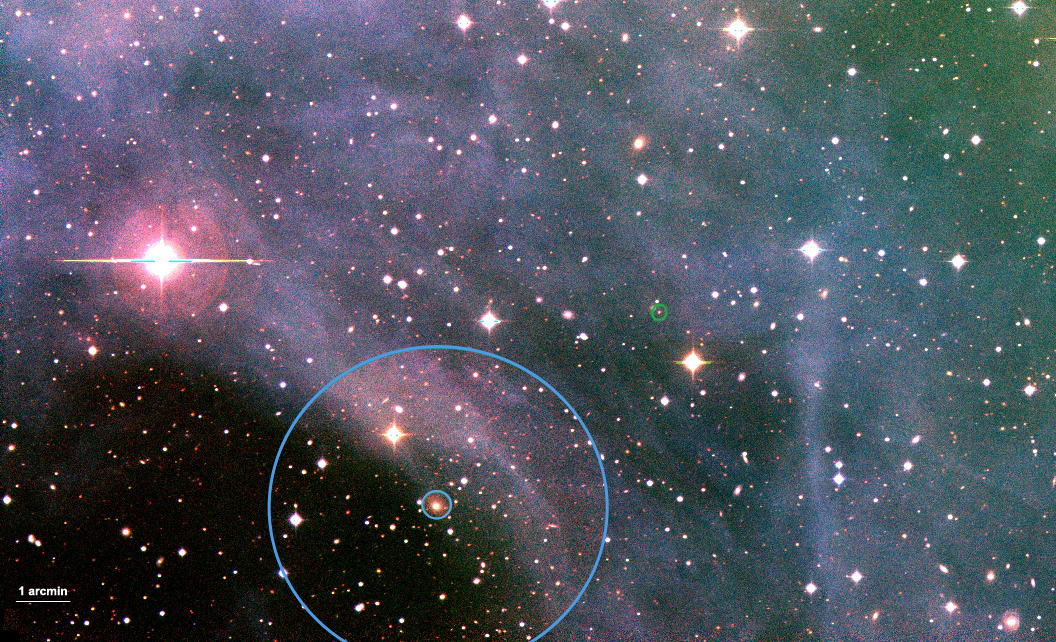}
\caption{ RBG image of the cluster PSZ1 G206.52$-$26.37, obtained with
  the WFC/INT. The green circle marks the \Planck\ PSZ1 position.  Blue circles
  show the location of the proposed cluster and the BCG. }
\label{fig:psz1-682}
\end{figure}

\paragraph{PSZ1 G206.52$-$26.37} The optical image (WFC/INT) shows strong 
extinction in the region, which might affect the photometric redshift estimate
(see Fig.~\ref{fig:psz1-682}). The proposed optical counterpart has a
photometric redshift of $\zphot = 0.36$. The BCG is $5\parcm2$ from the
\Planck\ position, and its spectroscopic redshift, obtained with ACAM/WHT, is
$z=0.391$.

\paragraph{PSZ1 G213.27$+$78.35} Two counterparts are identified for this SZ
candidate. The cluster reported in Table~\ref{tab:inpsz1} corresponds to
GMBCG J179.84162$+$26.45111 \citep{hao2010}, with photometric redshift
$\zphot=0.287$. This cluster is also described in \citet{Liu2014}, but the
quoted photometric redshift was $\zphot=0.316$. Here, we provide the
spectroscopic redshift of the cluster, using the information contained in the
SDSS DR10 database\footnote{SkyServer DR10:
  \url{http://skyserver.sdss.org/dr10/}.} for the BCG galaxy: $z=0.267$.  The
BCG of this cluster is $0\parcm78$ from the \Planck\ position.
Using the same tool, we also found a second galaxy cluster in the same
region. The BCG of this second cluster is at ($11^{\rm h}\ 59\arcm\ 20\parcs2$,
$26\deg\ 25\arcm\ 45\parcs1$), with a spectroscopic redshift of $z=0.139$. We
identified 28 spectroscopic members for this second cluster within $20\arcm$ of
the BCG.


\paragraph{PSZ1 G249.01$+$73.75} The optical counterpart identified in 
Table~\ref{tab:inpsz1} corresponds to a low-redshift cluster with two
spectroscopically confirmed members at $z=0.156$ using SDSS data. There are
several red galaxies with similar brightness around ($11^{\rm
  h}\ 57\arcm\ 05\parcs6$, $16\deg\ 57\arcm\ 35\parcs4$) with consistent
photometric redshifts. We provide this centroid as the coordinates for this
cluster. The apparent BCG is located at ($11^{\rm h}\ 56\arcm\ 59\parcs6$,
$16\deg\ 58\arcm\ 33\parcs1$). This target is identied in NED as photometric
cluster WHL J115659.6$+$165833 with $\zphot = 0.16$ \citep{WHL2012}.
There is another cluster in the region, WHL J115723.4$+$165903, $9\parcm5$ from
the \Planck\ position.  This cluster has $\zphot = 0.336$ \citep{wen2009}, and
here we spectroscopically confirm two members at $z=0.358$.

\subsection{Non-detections from the PSZ1 catalogue}
\label{sec:nondetections}

In five cases, our optical data are not sufficient to identify a counterpart of
the \Planck\ SZ source. In these cases, the most plausible explanations are
either that there is no optical counterpart (i.e. a false SZ detection due to
contamination by Galactic gas and dust emissions), or that the counterpart is at
high redshift and our imaging is not deep enough. The first four cases were
flagged as quality class 2 in the PSZ1 validation process
\citep{planck2013-p05a} (meaning that they fulfil good-quality criteria for the
SZ detection and for the associations and/or counterparts in ancillary data),
while the last case is quality class 3 (low-reliability cluster
candidate). We discuss these five cases in detail.

\begin{figure}[ht!]
\centering \includegraphics[width=\columnwidth]{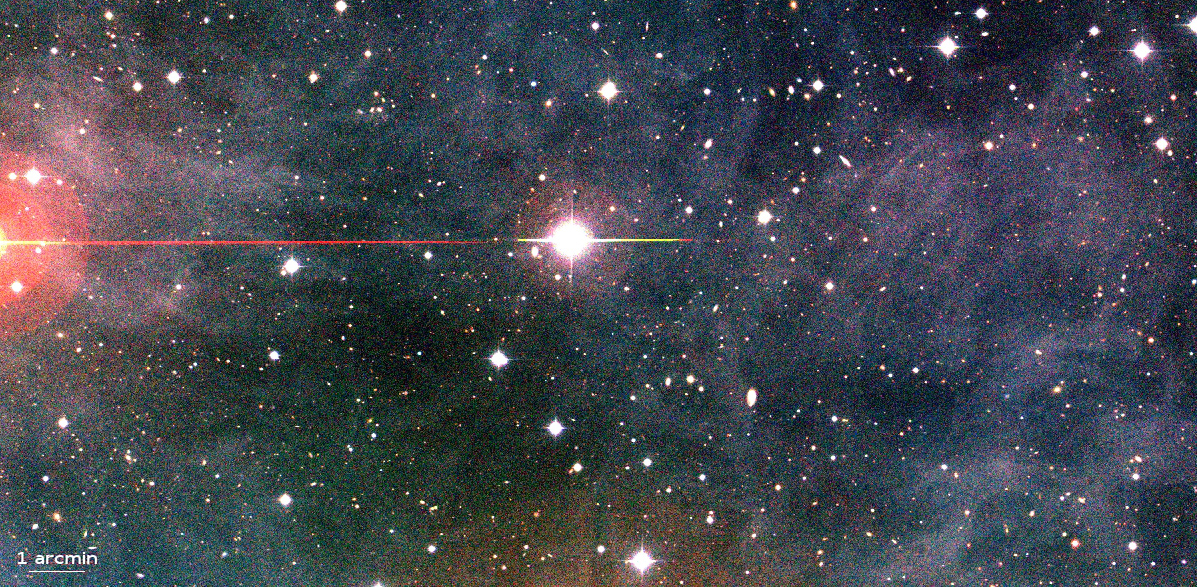}
\caption{ RGB image of PSZ1 G105.91$-$38.39 obtained with the WFC/INT,
  centred around the \Planck\ nominal coordinates for this target. This is an
  example of a \Planck\ SZ source without a clear optical counterpart, probably
  due to the impact of extinction from Galactic cirrus and the bright star near
  the image centre. }
\label{fig:psz1-380} 
\end{figure}

\paragraph{PSZ1 G105.91$-$38.39} There is no clear optical counterpart for this
SZ source. Our imaging data were obtained with WFC/INT down to $i' =25.6$. The
optical image shows several Galactic cirrus features and is strongly affected by
extinction and a bright star near the centre (see Fig.~\ref{fig:psz1-380}). We
found two overdensities of red objects. The first, $9\arcm$ from the
\Planck\ position, is unlikely to be associated with the SZ detection.  The
second is associated with WHL J235339.7$+$223928, a possible galaxy cluster at a
$\zphot=0.40$, $6\arcm$ from the \Planck\ position.  However, long-slit
observations with ACAM/WHT rule out this second case as a galaxy cluster because
the red galaxies have very different redshifts. We note that this target has
quality class 2 in the PSZ1 catalogue \citep{planck2013-p05a}, and it is
confirmed as a cluster by the Arcminute Microkelvin Imager (AMI)
\citep{AMI2014}.

\paragraph{PSZ1 G115.34$-$54.89} No counterpart is identified in our WFC/INT
imaging data down to $i' = 25$. This target was also observed with the RTT150
\citep{planck2014-XXVI} and no cluster counterpart was found. A bright star
close to the \Planck\ centre might affect the optical identification.

\paragraph{PSZ1 G127.55+20.84}  No counterpart is identified in our WFC/INT
imaging data down to $i' = 24.8$. The optical image shows several Galactic
cirrus features and is strongly affected by extinction.

\paragraph{PSZ1 G167.43$-$38.04} No counterpart is identified in our WFC/INT
imaging data down to $i' = 24.8$. This target was also discussed in
\citet{planck2014-XXVI}, with the same conclusion. The optical image shows
Galactic cirrus contamination.

\paragraph{PSZ1 G194.68$-$49.73} This target was observed with the RTT150 
\citep{planck2014-XXVI}, and no counterpart was found. We obtained images with
the WFC/INT and found a possible counterpart that we observed with OSIRIS/GTC.
We confirmed two isolated red galaxies with the same spectroscopic redshift of
$z=0.529$, at $3^{\rm h}\ 25\arcm\ 35\parcs0$, $-9\deg\ 36\arcm\ 10\parcs3$),
$\sim 4\arcm$ from the \Planck\ position. However, there are no additional red
objects in the surroundings with the same colours or redshifts. Therefore we do
not quote this case as a detection, although these two objects might form a
protocluster.

%

\subsection{Notes on individual detections outside the PSZ1 catalogue}

\begin{figure}[ht!]
\centering
\includegraphics[width=\columnwidth]{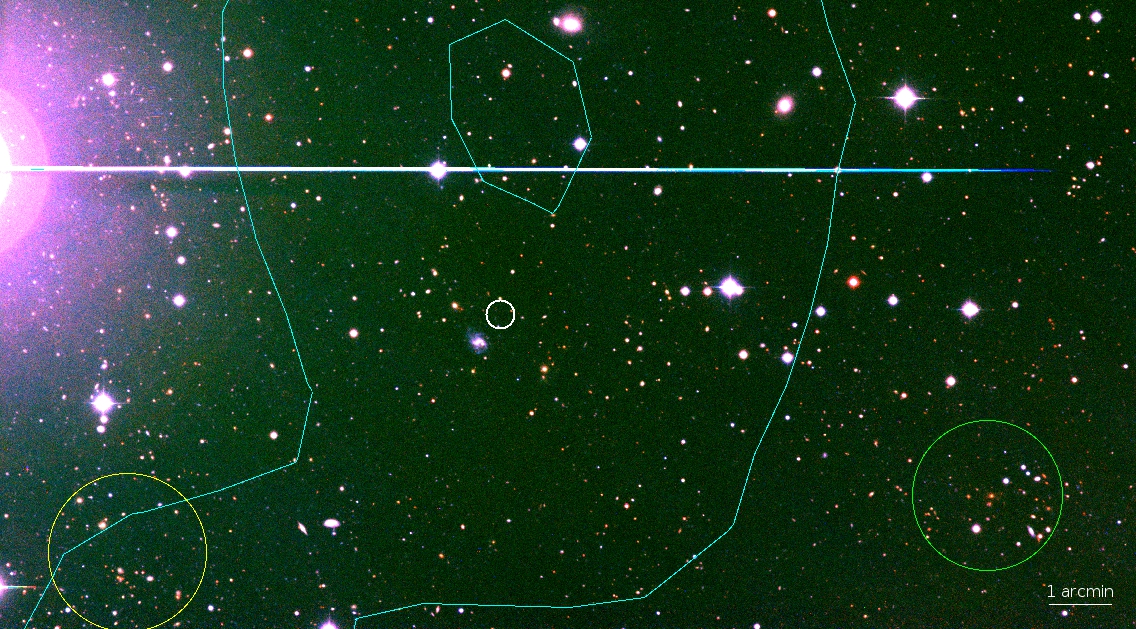}
\caption{RGB image of PLCK G128.38$+$71.18 obtained with the
  WFC/INT. The small white circle shows the \Planck\ position.  The yellow
  circle indicates our proposed counterpart, WHL J124204.5$+$454439, at
  a redshift of $z=0.3367$.  The green circle shows another cluster that we have
  spectroscopically confirmed, at $z=0.5775$. The light blue contours correspond
  to the $2.5\times 10^{-6}$ and $4\times 10^{-6}$ levels of the Compton $y$-map
  in this area \citep{planck2013-p05b}. See text for details. }
\label{fig:id1285}
\end{figure}

\paragraph{PLCK G128.38$+$71.18} This is a complicated case with four possible
counterparts, two of which we confirmed spectroscopically as clusters.
Figure~\ref{fig:id1285} shows our $20\arcm \times 10\arcm$ RGB image obtained
with the WFC/INT.
The counterpart listed in Table~\ref{tab:nopsz1} corresponds to a system $\sim
7\parcm3$ from the \Planck\ SZ position with 14 galaxies spectroscopically
confirmed with DOLORES/TNG and a BCG $z=0.3367$. This cluster is identified in
NED as WHL J124204.5$+$454439 \citep{WHL2012}.
Another cluster is $\sim 8\parcm7$ from the \Planck\ position, with BCG
(redshift 0.5775) located at ($12^{\rm h}\ 40\arcm\ 34\parcs8$,
$+45\deg\ 47\arcm\ 31\parcs8$), and ten spectroscopically confirmed members with
DOLORES/TNG.
We found two more possible clusters.  One is listed in NED as WHL
J124151.2$+$455301 and $\zphot=0.4764$ \citep{WHL2012}, but our spectroscopic
observations with the MOS at DOLORES/TNG are not conclusive for this target. Of
the 20 potential members of the cluster, we have only three almost coincident
spectroscopic redshifts around $z=0.496$. Thus, we cannot confirm this
overdensity as a real cluster, which in case of confirmation would correspond to
a poor group.
The fourth overdensity of red objects in the region is associated with NSCS
J124127$+$455030 in the NED database \citep{NSCS}. However, our MOS observations
with DOLORES/TNG for this clump do not find coincident redshifts for the
potential members, therefore this candidate does not seem to be a real cluster.

\paragraph{PLCK G134.25$-$44.24} There are no clear counterparts within $5\arcm$
of the \Planck\ coordinates, but there are two possible counterparts farther
away. Taking into account the spatial distribution of the signal in the Compton
$y$-map, we consider as the most probable candidate a low-redshift cluster
located at $\sim 8\parcm9$ from the \Planck\ position. This cluster has 4
spectroscopically confirmed members at $< z > = 0.232$, and more than 25
photometric members. In addition, there is a closer second candidate at
($01^{\rm h}\ 25\arcm\ 40\parcs89$, $+17\deg\ 49\arcm\ 18\parcs9$) with 6
spectroscopic members at $< z > = 0.427$. Further observations are needed to
determine which cluster provides the larger contribution to the SZ signal.

\paragraph{PLCK G225.44$+$51.89} This is a very poor system $3\parcm7$ from 
the \Planck\ position, with two galaxies spectroscopically confirmed at
$z=0.297$.  There is no clear BCG, so we report the average central coordinate
of all photometrically selected members.  There is another group of galaxies
$5\parcm5$ to the south of the \Planck\ position, with ten photometric members
with $\zphot=0.082$. There is no evident BCG for this group, but there are seven
spectroscopic determinations in SDSS, with a mean redshift of $z=0.084$.

\section{Conclusions}
\label{sec:conclusions}

This article is a companion paper to the \Planck\ catalogue of SZ sources (PSZ1)
published in \citet{planck2013-p05a}. It contains the results of approximately
three years of observations with telescopes at the Canary Islands observatories
(IAC80, NOT, INT, TNG, WHT, and GTC), as part of the general optical follow-up
programme undertaken by the \Planck\ collaboration.

In total, 78 SZ sources were discussed here. Deep-imaging observations were
obtained for most; in many cases, spectroscopic observations were carried out
either in long-slit or multi-object spectroscopic modes.  We found optical
counterparts for 73 of the 78 candidates. The confirmations include 53
spectroscopic redshift determinations, 20 of them obtained with a multi-object
spectroscopic mode. Twenty-seven of these sources were not included in the first
\Planck\ SZ source catalogue (PSZ1) and are published here for the first time.

Table~\ref{tab:inpsz1} contains five cases (out of 67) in which more than one
possible cluster counterpart is identified along the line of sight of the SZ
signal. Similar projection effects were also discussed in earlier works with a
subsample of \Planck\ clusters observed with XMM-Newton \citep{planck2012-IV} or
with the RTT150 \citep{planck2014-XXVI}.  However, as discussed in detail in the
RTT150 paper, the sub-sample of \Planck\ SZ sources included in this work is not
statistically representative in any sense. Sources were selected from different
versions of the \Planck\ SZ source catalogue during a two-year period, and after
the PSZ1 catalogue was completed, observations were focused on the clusters
 without optical counterparts. It is therefore not possible to infer the
statistical properties of the full PSZ1 sample (e.g. biases, fraction of false
sources, fraction of sources with projections) using only the targets
listed in Table~\ref{tab:inpsz1}. Any statistical analysis needs
to be based on
the full PSZ1 sample.

After this work, the current status of the PSZ1 validation, including all
confirmed clusters, is summarized in Table~10 of \citet{planck2014-a36}.

We emphasize the importance of using the Compton $y$-map \citep{planck2013-p05b}
as a tool to confirm the optical counterparts of some candidates. Here we
discussed several cases (e.g. PSZ1 G046.98$+$66.62, PSZ1 G153.87$+$41.05) of
clusters with anomalously large angular separations with respect to the nominal
\Planck\ SZ centre, but whose locations match the peak of the SZ emission in the
$y$-map very well.

On-going (and future) follow-up programmes, some of them running in the
facilites described in this paper, are expected to be able to complete the optical
identification of \Planck\ cluster candidates within a few years.

\begin{acknowledgements}
The Planck Collaboration acknowledges the support of: ESA; CNES, and
CNRS/INSU-IN2P3-INP (France); ASI, CNR, and INAF (Italy); NASA and DoE (USA);
STFC and UKSA (UK); CSIC, MINECO, JA and RES (Spain); Tekes, AoF, and CSC
(Finland); DLR and MPG (Germany); CSA (Canada); DTU Space (Denmark); SER/SSO
(Switzerland); RCN (Norway); SFI (Ireland); FCT/MCTES (Portugal); ERC and PRACE
(EU). A description of the Planck Collaboration and a list of its members,
indicating which technical or scientific activities they have been involved in,
can be found at
\href{http://www.cosmos.esa.int/web/planck/planck-collaboration}{\texttt{http://www.cosmos.esa.int/web/planck/planck-collaboration}}.
This article is based on observations made with a) the Gran Telescopio Canarias
(GTC), installed in the Spanish Observatorio del Roque de los Muchachos (ORM) of
the Instituto de Astrof\'{\i}sica de Canarias (IAC), in the island of La Palma;
b) the Isaac Newton Telescope and the William Herschel Telescope operated on the
island of La Palma by the ISAAC Newton Group of Telescopes in the Spanish ORM of
the IAC; c) the italian Telescopio Nazionale Galileo (TNG) operated on the
island of La Palma by the Fundaci\'on Galileo Galilei of the INAF (Istituto
Nazionale di Astrofisica) at the Spanish ORM of the IAC; d) the Nordic Optical
Telescope, operated on the island of La Palma jointly by Denmark, Finland,
Iceland, Norway, and Sweden, in the Spanish ORM of the IAC; and e) the IAC80
telescope operated on the island of Tenerife by the IAC in the Spanish
Observatorio del Teide.
This research has been carried out with telescope time awarded by the CCI
International Time Programme at the Canary Islands Observatories (programmes
ITP12-2 and ITP13-8).
This research has made use of the following databases: the NED database,
operated by the Jet Propulsion Laboratory, California Institute of Technology,
under contract with NASA; SIMBAD, operated at CDS, Strasbourg, France; the
SZ-Cluster Database operated by the Integrated Data and Operation Center (IDOC)
at the IAS under contract with CNES and CNRS; and the SDSS.
Funding for the Sloan Digital Sky Survey (SDSS) has been provided by the Alfred
P. Sloan Foundation, the Participating Institutions, the National Aeronautics
and Space Administration, the National Science Foundation, the U.S. Department
of Energy, the Japanese Monbukagakusho, and the Max Planck Society.
AS, RB, HL, and JARM acknowledge financial support from the Spanish Ministry of
Economy and Competitiveness (MINECO) under the 2011 Severo Ochoa Program MINECO
SEV-2011-0187, and the Consolider-Ingenio project CSD2010-00064 (EPI: Exploring
the Physics of Inflation).
\end{acknowledgements}

\bibliographystyle{aa}
\bibliography{Planck_bib,pip23}

\end{document}